\documentstyle[12pt]{article}
 
\def\b{\beta}
\def\c{\chi}
\def\d{\delta}

\def\g{\gamma}

\def\j{\psi}
\def\k{\kappa}     
\def\l{\lambda}
\def\m{\mu}
\def\n{\nu}
\def\o{\omega}
\def\p{\pi}       

\def\x{\xi}

\def\J{\Psi}

\def\O{\Omega}

\def\cb{\overline{\c}}

\def\jb{\overline{\j}}
\def\ob{\overline{\o}}

\def\xib{\overline{\x}}

\def\Psb{\overline{\J}}

\def\pst{\j}
\def\psbt{\jb}
\def\ct{\c}
\def\cbt{\cb}

\def\Vt{V}
\def\psb{\jb}
\def\psib{\jb}

\def\mb{\overline{m}}
\def\mh{\hat{\m}}
\def\ph{\hat{p}}
\def\nh{\hat{\n}}

\def\del{\partial}
 
\def\ds{\del\!\!\!/\,}
\def\Ds{D\!\!\!\!/\,}
\def\ssl{s\!\!\!/}
\newcommand{\ncm}{\newcommand}
\ncm{\rencm}{\renewcommand}
\ncm{\dsp}{\displaystyle}
\ncm{\nn}{\nonumber\\}
\ncm{\nit}{\noindent}
\ncm{\av}[1]{\mbox{$\langle #1 \rangle$}}
\ncm{\avc}[1]{\mbox{$\langle #1 \rangle_{\psi}$}}
\ncm{\half}{\mbox{{\small $\frac{1}{2}$}} }
\ncm{\quart}{\mbox{{\small $\frac{1}{4}$}} }
\ncm{\tq}{\mbox{{\small $\frac{3}{4}$}} }
\ncm{\third}{\mbox{{\small $\frac{1}{3}$}} }
\ncm{\sixth}{\mbox{{\small $\frac{1}{6}$}} }
\ncm{\eigth}{\mbox{{\small $\frac{1}{8}$}} }
\ncm{\thrhalf}{\mbox{{\small $\frac{3}{2}$}} }
\ncm{\thrfor}{\mbox{{\small $\frac{3}{4}$}} }
\ncm{\twothi}{\mbox{{\small $\frac{2}{3}$}} }
\ncm{\fivtwo}{\mbox{{\small $\frac{5}{2}$}} }
\ncm{\ninhalf}{\mbox{{\small $\frac{9}{2}$}} }
\ncm{\ninth}{\mbox{{\small $\frac{1}{9}$}} }
\ncm{\nist}{\mbox{{\small $\frac{9}{16}$}} }
\ncm{\RE}{\mbox{\,Re\,}}
\ncm{\IM}{\mbox{\,Im\,}}
\ncm{\Tr}{\mbox{\,tr\,}}
\ncm{\Det}{\mbox{Det}}
\ncm{\diag}{\mbox{diag}}
\ncm{\acos}{\mbox{acos}}
\ncm{\ra}{\rightarrow}
\ncm{\la}{\leftarrow}
\ncm{\dg}{\dagger}
\ncm{\ha}{\hat{a}}
\ncm{\hP}{\hat{P}}
\ncm{\sL}{\sqrt{\Lambda}}
\ncm{\lb}{\overline{\lambda}}
\ncm{\aldot}{\mbox{$\dot{\alpha}$}}
\ncm{\xdot}{\mbox{$\dot{x}$}}
\ncm{\dotf}{\mbox{$\dot{\phi}$}}
\ncm{\dfo}{\mbox{$\partial_{\phi_0}$}}
\ncm{\aplt}{ \mbox{}_{\textstyle \sim}^{\textstyle < }     }
\ncm{\apgt}{ \mbox{}_{\textstyle \sim}^{\textstyle > }     }
\ncm{\Oa}{\mbox{$\mbox{O}(a)$}}
\ncm{\Sp}{\hspace{1.0cm}}
\ncm{\bibi}{\bibitem}
\def\be{\begin{equation}}
\def\ee{\end{equation}}
\def\bea{\begin{eqnarray}}
\def\eea{\end{eqnarray}}
\ncm{\beba}[1]{\be\begin{array}{#1}}
\def\eeea{\end{array} \ee}
\def\ba{\begin{array}}
\def\ea{\end{array}}
\ncm{\fig}[2]{\begin{figure}[ppp]\caption{\label{#1} #2}\end{figure}}
\rencm{\theequation}{\arabic{section}.\arabic{equation}}
\ncm{\sect}[1]{\section{#1}\setcounter{equation}{0}}
\ncm{\append}{  \setcounter{section}{1}\setcounter{equation}{0}
          \section*{Appendix}
   \rencm{\theequation}{\Alph{section}.\arabic{equation}}  }
\ncm{\appendA}{  \setcounter{section}{1}\setcounter{equation}{0}
          \section*{Appendix A}
   \rencm{\theequation}{\Alph{section}.\arabic{equation}}  }
\ncm{\appendB}{  \setcounter{section}{2}\setcounter{equation}{0}
          \section*{Appendix B}
   \rencm{\theequation}{\Alph{section}.\arabic{equation}}  }
\ncm{\front}[5]{
\begin{titlepage}
\hbox{}\vspace{1.cm}
\noindent {#1} \hfill {#2} \\
\begin{center}
\vspace{1\baselineskip}
{\Large\bf  #3  } \\
\vspace{1.3\baselineskip}
\vspace{1.0\baselineskip}
 #4\\
\vspace{1.0\baselineskip}
$\mbox{}^a$Washington University, Deparment of Physics,
St.~Louis,~MO~63130,~USA \\
$\mbox{}^b$University of California at San Diego,
Department of Physics 0319, La~Jolla,~CA~92093,~USA
 
\end{center}
\vfill
{\bf Abstract}\\
 #5
\end{titlepage} }
 
\begin{document}
\front{September 1993} {UCSD/PTH 93-28, Wash.~U.~HEP/93-60\\ 
                   \hbox{} \hfill  hep-lat 9309015}
{Investigation of the domain wall fermion approach to chiral gauge theories
                         on the lattice }
{ Maarten F.L. Golterman$^{a,}$\footnote{e-mail: maarten@wuphys.wustl.edu}, 
Karl Jansen$^{b,}$\footnote{e-mail: jansen@higgs.ucsd.edu,\\ 
\Sp Address after October 1: DESY, Notkestrasse 85, 22603 Hamburg 52, Germany}, 
Donald N. Petcher$^{a,}$\footnote{e-mail: 
petcher@moriah.covenant.edu,\\ \Sp present address: Dept. of Phys.,
Covenant College, Lookout Mountain, GA 30750, USA } \\ 
and Jeroen C. Vink$^{b,}$\footnote{e-mail: vink@yukawa.ucsd.edu}}
{We investigate a recent proposal to construct chiral gauge theories
on the lattice using domain wall fermions. 
We restrict ourselves to the finite volume case, in which two 
domain walls are present, with modes of opposite chirality on each of them.
We couple the chiral fermions
on only one of the domain walls to a gauge field. In order to preserve
gauge invariance, we have to add a scalar field, which
gives rise to additional light mirror fermion and scalar modes. We argue
that in an anomaly free model these extra modes would decouple if our
model possesses a so-called strong coupling symmetric phase. However,
our numerical results indicate that such a phase most probably does not
exist.
}

\sect{Introduction}

The lattice provides a first-principles
regularization of quantum field theories, which allows us to explore the 
nonperturbative properties of a model and for vectorlike theories, 
such as QCD, it has proven to be very successful.  Since the full 
Standard Model is a chiral gauge theory, it is natural to attempt 
a construction of chiral gauge models on the lattice as well.

As is well known, however, on the lattice one is confronted with 
``species doubling'' \cite{TRIANG,DOUB}, i.e. the phenomenon that
a single Weyl fermion field on the lattice leads to 
an equal number of left and right handed fermions in the continuum
limit.  When coupled to a gauge field, all doublers transform in the
same representation of the gauge group which
prevents an easy construction of chiral gauge theories.
For nonchiral models there are two well tested ways of dealing with the 
species doublers: 
they can be decoupled with a momentum dependent mass term as in
Wilson's method, or they can be used as physical degrees of freedom as in
the staggered fermion method.
However, since these methods violate chiral symmetry, a
straightforward extension to chiral gauge theories would clash with
gauge invariance.  Proposals for chiral gauge theories
on the lattice include generalizations of Wilson's method 
[3--6]
and of the staggered method \cite{Sm88}. 
There are proposals that try to avoid coupling the doublers \cite{ZA}, 
and approaches that start from a gauge-fixed
continuum action \cite{ROME,BK}. There are also proposals for more
radical departures from the usual lattice fermion prescriptions
[11--13].
For a recent review, see ref. \cite{REV}.

The domain wall fermion approach suggested in ref. \cite{Kapl92}
falls into the last group and has attracted a lot of attention recently
[15--22].
In the domain wall model
an extra  dimension is added to our four dimensional world. In this
five dimensional world the model is vectorlike and the fermion doublers
can be removed using Wilson's method without breaking gauge invariance. 
The reduction to a four
dimensional world with a chiral fermion is made by giving the fermions
a mass term which flips sign across a four dimensional domain wall.
It has been shown \cite{Kapl92} that the lattice Wilson-Dirac operator 
with such a mass term has a chiral zeromode, which is bound to the 
domain wall.  This fermion remains massless and localized at the domain
wall for (four-)momenta below a critical momentum \cite{Kapl92,Jans92,JaSc92}.
On a finite lattice the (periodic) boundary conditions lead to a second
anti-domain wall with a chiral fermion of opposite handedness.

Every lattice model for a chiral gauge theory has to produce the 
appropriate anomaly structure of the target continuum theory.
The domain wall model has the potential to solve this problem elegantly 
with the help of the extra dimension.
The starting five dimensional model is vectorlike and hence the gauge 
current $J_\m$ is anomaly free, $\sum_{\m=1}^{5} \del_\m J_\m = 0$.
However, the four dimensional
current restricted to the domain wall is clearly not
conserved $\sum_{\m=1}^{4} \del_\m 
J_\m = -\del_{5} J_{5}$, and its divergence reproduces 
the expected anomaly when computed for weak external gauge fields
\cite{Kapl92,Jans92,GoJa92}.  
$J_5$ takes the form of a Goldstone-Wilczek
current with a nonzero derivative across the domain wall, as was 
demonstrated some time ago in the continuum in ref. \cite{CaHa85}.
On a finite lattice this Goldstone-Wilczek current transports charge
from the domain wall to the anti-domain wall, ensuring charge conservation
in the five dimensional theory. 
The same mechanism should also yield the correct four dimensional
global anomaly structure. 

In the work referred to above the domain wall fermions are coupled to
fixed smooth external gauge fields. Here it is not important that the
chiral fermions at both the domain and anti-domain
walls couple to the gauge field,
because we can single out one of the domain walls by hand. With
dynamical gauge fields, however, the crucial requirement is that only
a single domain wall fermion couples to the gauge field. If this can be 
achieved the second domain wall can be ignored and we are left with an 
interacting chiral fermion in a four dimensional world located at the
domain wall, assuming that the fermions are in an anomaly free 
representation.

In the original proposal it was hoped that the communication between
the two domain walls could be prevented by modifying the gauge
interactions in the fifth dimension. However, it seems likely that
this approach does not lead to the desired decoupling
(see also ref. \cite{Kort93}) and here we follow instead the suggestion 
made in ref. \cite{Kapl93}.
In this  approach the gauge fields are coupled only in a restricted 
region around one of the domain walls, where the size of this region, 
which we will call the waveguide, should be at least as large as 
the support of the wave function of the domain wall zeromode.
However, as will be discussed in much more detail
below, the requirement of gauge invariance leads to the introduction of
an extra scalar field at the boundaries of the waveguide. 
This scalar field screens the gauge charge of the fermions at the 
waveguide boundary and allows for interactions between these charged 
fermions and the neutral ones outside the waveguide. This leads to  
Yukawa couplings located at the waveguide boundary which give rise 
to additional light fermion modes at the waveguide boundary. This is
most easily seen for zero Yukawa coupling, because then the  waveguide
region decouples from the  anti-domain wall region and five dimensional
charge conservation is now ensured by the new zeromodes at the boundary.  

The aim is then to decouple the fermion at the waveguide boundary and
maintain at the same time the chiral zeromode at the domain wall. 
Because we have introduced a scalar field coupled to the fermions through
a Yukawa interaction, we may hope for a rich phase structure of the
model, similar  to that found in other two and four
dimensional Yukawa models on the
lattice. In particular, one expects that one can
drive the system into a symmetric phase, with vanishing scalar field vacuum
expectation value $v$, and a spontaneously broken phase, with $v>0$.
For small values of the Yukawa coupling one then expects the fermions 
at the waveguide to follow the perturbative relation $m_F \propto v$ 
(with $v=0$ in the symmetric phase). This means that these
fermions remain light and appear in the low energy spectrum. 
However, at large values of the Yukawa coupling, the interaction of the
fermion and scalar fields might become so strong that only a 
bound state fermion exists
with a mass of the order of the cutoff. Such a strong coupling behavior
has been established in various Yukawa models on the lattice 
[25--27].
If the four dimensional model is anomaly free and if we choose the
waveguide boundary with the scalar field far enough from the domain
wall, we could hope to take the waveguide fermions into a strong coupling 
symmetric phase, without affecting the chiral mode at the domain wall.
Then the fermions at the waveguide boundary would decouple from the 
low energy physics, leaving only the chiral zeromodes at the domain 
wall coupled to the gauge field.

The crucial question we will investigate in this paper is therefore
whether such a strong symmetric phase exists in the domain wall model.
We will present evidence based on
analytical considerations and numerical results, which leads
us to conclude that an appropriate
strong coupling phase  most probably does not occur.
 
The paper is organized as follows.
In sect. 2 we review free domain wall fermions, discuss the coupling to
the gauge field and the  need to introduce the extra scalar field.
We close the section with a  sketch of the  phase diagram we would hope to 
find for our model.
In sect. 3 we rewrite the fermion action in a mirror-fermion form, such
that we can distinguish the light modes from the heavy ones.
In the next section we present results for fermion masses, concentrating 
on the results for the boundary fermion. In sect. 5 we continue our search
for a strong coupling phase using the eigenvalue spectra of the fermion 
matrix in a simplified model, in which all heavy modes are discarded.
Sect. 6 contains a brief discussion of alternative ways to couple the
gauge field and in sect. 7 we present our conclusions.

\sect{Domain wall fermions coupled to gauge fields}
\subsection{Resum\'e of free domain wall fermions}

Let us start our discussion with a short resum\'e of free domain wall
fermions.
Consider an odd dimensional lattice of size $L^dL_s$, 
with $d=2n$, $L_s$ the extent in the extra dimension and lattice sites 
labeled by $(x,s)$, $(x\equiv (x_1,\cdots,x_d))$. 
The action for free domain wall fermions \cite{Kapl92} can be written as, 
\bea
  S_{\J} & = & \sum_s \left( \sum_{xy}
   \Psb_x^s(\ds_{xy} -w_{xy} + m^s\d_{x,y})\Psi_y^s \right. \nn
    & - &  \half\sum_x[  \left.
        \Psb_x^s(r-\g_5)\Psi_x^{s+1} + \Psb_x^{s+1}(r+\g_5)\Psi_x^s
        -2r\Psb_x^s\Psi_x^s] \right),
                   \label{SF1}
\eea
where $\ds$ and $w$ are the Dirac operator and Wilson term with Wilson
parameter $r$ on the even
$d$ dimensional lattice,
\bea
  \ds_{xy}  & = & \sum_{\m=1}^d \half \g_{\m}
                         [\d_{x+\mh,y} - \d_{x-\mh,y}],  \nn
  w_{xy}   & = & \sum_{\m=1}^d \half 
                   r[\d_{x+\mh,y} + \d_{x-\mh,y} -2\d_{x,y}]. 
\eea
The $\d_{x,y}$ is the Kronecker delta and we use lattice units $a=1$.
We shall choose the Wilson parameter $r=1$ for convenience and
for the domain wall mass, denoted by $m^s$,  we choose a periodic 
step function of the form (with $L_s$ even),
\beba{lcrl}
  m^s & = & -m_0 & \;\;\; s=2,\cdots,L_s/2, \nn
  m^s & = &   0  & \;\;\; s=1,L_s/2+1, \nn
  m^s & = & +m_0 & \;\;\; s=L_s/2+2,\cdots,L_s. 
\eeea
With periodic boundary conditions in $s$, the emergence of an
anti-domain wall is inevitable.
It will often be convenient to think of $s$ as a flavor label, rather
than an extra space-time coordinate.

This model posesses
two chiral zeromodes with the property that the mode bound to the
domain wall at $s=1$ is left handed 
($\gamma_5=-1$) and the mode bound to the other
domain wall at $s=L_s/2+1$ is right handed 
($\gamma_5=+1$) \cite{Kapl92}. The wave functions for both modes
have the form of plane waves in the $2n$-dimensional space and decay
exponentially in $s$, away from the domain walls. These chiral zeromodes
exist for plane wave momenta below some critical momentum
$p_c$ which depends on the ratio $m_0/r$. 
For different values of $m_0/r$ the zeromode spectrum can change
substantially. For $0<m_0/r < 2$ one has only one chiral zeromode at each
domain wall.
For increasing values of $m_0/r$ this zeromode becomes less
localized and disappears at $m_0/r=2$.
At this point  new zeromodes with opposite chirality are provided by
the species doublers which are located at
different corners of the Brillouin zone \cite{JaSc92,GoJa92}. 
Throughout the paper we will take $m_0/r \approx 1$ and hence we will have 
only one chiral zeromode at the domain wall with exponentially small overlap
with the zeromode at the anti-domain wall.
In this case, chiral modes exist for momenta $p$ below a critical momentum, 
$|{\hat p}|<p_c$, with ${\hat p}^2=2\sum_\mu(1-\cos{(p_\mu)})$ and 
$p_c^2=4-2m_0/r$.  Note that ${\hat p}^2\approx p^2$ for small momenta.

\subsection{Coupling to gauge fields}

Since the left and right handed zeromode 
components of the fermion field now are 
separated in $s$ space, one can attempt to couple these two components in
different ways to a gauge field. If we succeed in coupling only one of 
the two zeromodes to a gauge field, we can hope to use this in order to
construct a chiral gauge theory on the lattice.  
This appears to be impossible if one also insists that gauge invariance
is maintained. However, if we do not worry about gauge invariance for
the moment, we can couple the right handed mode to a gauge field, by
replacing the free ($2n$-dimensional) Dirac operator and Wilson term 
by the gauge invariant ones, but only for a restricted number of 
$s$-slices around the right-handed domain wall, cf. \cite{Kapl93}. 
In this way the gauge field is confined within a ``waveguide'' 
around the domain wall, and interactions
with the opposite chirality mode at the anti-domain wall are
exponentially suppressed with $L_s$.

We take the same gauge field on all $s$-slices inside the waveguide, 
which is natural if one thinks of $s$ as a flavor label, 
and define gauge transformations on the fermion field as 
\beba{rclrcll}

\Psi_x^s & \ra & g_x\Psi_x^s,&  \Psb_x^s & \ra & \Psb_x^s g_x^{\dg} & \;\; 
               s \in WG , \nn 
\Psi_x^s & \ra & \Psi_x^s,& \Psb_x^s & \ra & \Psb_x^s & \;\; 
                 s \not\in WG ,   \label{LIF}
\eeea
\be  
           WG  =  \{s: s_0\leq s \leq s_0'\}     \label{WG}
\ee
with $g_x$ in a gauge group $G$. 
The detailed choice of the boundaries $s_0$
and $s_0'$ is not very important, provided they are sufficiently far
from the domain wall that the zeromode is exponentially small at
the waveguide boundary. For symmetry reasons we shall choose 
$s_0 = (L_s+2)/4+1$ and $s_0' = (3L_s+2)/4$, such that the right 
handed mode at $s=L_s/2+1$ is  located at the center of the
waveguide, see fig.\ 1.
With this choice we have to  take $L_s-2$ a multiple of four. 

Having made this division into a waveguide and its exterior, we note
that the model has a global $G\times G$ symmetry:
\beba{rclrcll}
\Psi_x^s & \ra & g\Psi_x^s, & \;\Psb_x^s& \ra & \Psb_x^s g^{\dg}, & \;\; 
                                  s \in WG,  \nn
\Psi_x^s & \ra & h\Psi_x^s,\; & \Psb_x^s & \ra & \Psb_x^s h^\dg,& 
           \;\; s \not\in WG.
                                        \label{GLOBAL}
\eeea
With our choice for
the position of the waveguide boundary, there is a symmetry
involving parity plus a reflection in the 
$s$-direction with respect to the plane $s=s_0-\half=L_s/4+1$,
\be
\Psi_x^s \ra \g_d\Psi_{Px}^{L_s/2+2-s},   \label{PARITY}
\ee
with $Px = (-x_1,\cdots,-x_{d-1},x_d)$ the parity transform of $x$.

It is clear that the hopping terms from $s_0-1$ to $s_0$ and from 
$s_0'$ to $s_0'+1$ break the  local gauge invariance of eq. (\ref{LIF}). 
However, this can be repaired
by putting in a scalar field $V$ at the boundary of the waveguide, or 
alternatively by interpreting the gauge 
field $g_x$ that appears in the action after performing a gauge 
transformation as a  St\"uckelberg field. 
 This leads to the gauge invariant action
\bea
 S_\J 
       & = & \sum_{s\in WG} \Psb^s(\Ds(U) -W(U) + m^s)\Psi^s
 + \sum_{s\not\in WG}  \Psb^s(\ds-w+ m^s)\Psi^s \nn
  & - & \sum_{s\not=s_0-1,s_0'}[\Psb^sP_L\Psi^{s+1}
                         + \Psb^{s+1}P_R\Psi^s ]+\sum_s \Psb^s\Psi^s
                    \label{SFUV}\\ 
 & - &  y(\Psb^{s_0-1}VP_L\Psi^{s_0}  +  \Psb^{s_0}V^{\dg}P_R\Psi^{s_0-1})
  -   y(\Psb^{s_0'}V^{\dg}P_L\Psi^{s_0'+1} + \Psb^{s_0'+1}VP_R\Psi^{s_0'}),
   \nonumber
\eea
where we have supplied the Yukawa term  with a coupling constant $y$.
Note that we take the same scalar field at both waveguide boundaries.
Since we have chosen $r=1$ we have written projectors in the hopping terms 
in $s$, $P_{R(L)} = \half(1 +(-)\g_5)$. $\Ds(U)$ and $W(U)$ are the 
usual gauge covariant Dirac operator and
Wilson term, whose explicit form is not important here, since we shall only
work with $U=1$ in this paper. The field $V_x \in G$ is the scalar
field, which can be thought of as a (radially frozen) 
Higgs field, and which transforms as
\be
         V_x \ra hV_xg_x^\dg.   \label{LIB}
\ee
The transformation given in eq. (\ref{PARITY}) remains a symmetry if 
$V$ transforms as
\be
	 V_x \ra V^\dg_{Px}.    \label{VPARITY}
\ee

Since gauge invariance is broken in the model without scalar field, we
add a mass term for the gauge boson, which on the
lattice takes the form $\k \sum_\mu \Tr(U_\mu + U_\mu^\dg)$, with $\k$ the
mass parameter in lattice units.
It takes the form of a hopping term for $V$ when this field is used to 
restore the gauge invariance of this mass term, 

\be
S_V 
    = -\k \sum_{x,\mu} \Tr( V_x U_{\m x} V_{x+\mh}^\dg + h.c.).
                          \label{SV}
\ee

Gauge invariance and the necessity to couple only the zeromode on one
of the domain walls to the gauge field has led to an action which
contains an additional scalar field. 
One might wonder whether there is 
a better way to introduce a gauge  field which couples to only
one of the domain wall zeromodes, but
avoids the extra scalar field. Unfortunately, this appears 
to be difficult, if not impossible in a model which contains
both domain walls,  as we shall argue in sect. 6. 
For a proposal in a different direction,  in which the anti-domain wall
is avoided by keeping $L_s$ strictly infinite, see ref. \cite{NaNe93}.

To get an idea about the physics of the model (\ref{SFUV}), we can start 
with $y=0$,  
in which case the scalar field is decoupled. However, now the
gauged and ungauged parts of the action have decoupled completely as
well, which implies that the two zeromodes on the domain walls are no 
longer balanced by each other. Therefore new zeromodes with opposite
chirality must emerge which will be bound to the waveguide boundary. 
As an illustration we have plotted in fig.\ 1
the four zeromodes computed for the smallest plane wave
momentum on a lattice with $d=2$, $L_s=50$ and $U=1$. 
 At $y=0$ (figure 1a) one recognizes the two expected massless 
modes at the domain walls,
but also two modes at the waveguide boundary. These modes are 
massless\footnote[1]{Of course, these modes are not exactly massless,
because of the exponentially suppressed mixing between the domain wall
and boundary modes.},
because there can be no overlap between the left and right handed components
across the waveguide boundary. 
For nonzero $y$ the two components can overlap and they form a Dirac state
with mass approximately equal to $y$, see fig.\ 1b,
where we took $V=1$.  One clearly sees how in this case the wave functions
which are peaked at the waveguide  boundary extend across this boundary. 
Note that the modes shown in fig.\ 1  are symmetric around the
waveguide boundary, in accordance with the symmetry given in
eq. (\ref{PARITY}).
The extra mirror modes  at the waveguide boundary will be further 
discussed in the next section.

\subsection{Conjectured phase diagram}

To arrive at a chiral model, both this additional fermion at the
waveguide boundary and the scalar field have to be decoupled. 
We first make the simplification of neglecting the gauge field dynamics
by replacing $U \ra 1$. This is reasonable, because we are interested in
the scaling region at small gauge coupling. There we can write
$U_{\m x} = \O_x U^L_{\m x}\O^\dg_{\m x+\mh}$ with  $U^L$ the gauge
field in the smooth Landau gauge. The $\O$ can be absorbed by a gauge
transformation on $\psi$, $\psb$ and $V$.
Since $U^L$ is now smooth and close to one, we  can treat this field 
in perturbation theory and put $U=1$ in our numerical computations.
Note that at large Yukawa couplings the scalar field dynamics cannot 
be computed in perturbation theory.

The light fermion modes are
well localized, which implies that the mode at the domain wall has only
an exponentially small overlap with the  scalar field at the waveguide
boundary. With $r=1$ the domain wall zeromode has a magnitude $\propto
e^{-m_0 L_s/4}$  at the waveguide boundary,
and for sufficiently large $L_s$ the effective Yukawa
coupling to the scalar field is exponentially suppressed, $ye^{-m_0L_s/4}$.
The new fermion mode, on
the other hand, is localized at the $s$-slice that carries the scalar
field and is coupled to it with strength $y$.  
Therefore we can use the freedom of adjusting the coupling constants 
$y$ and $\k$ to try to decouple the unwanted fields at the waveguide
boundary,  while  keeping $L_s$ sufficiently large as to ensure that
the physics of the zeromode at the domain wall will remain unaffected.

In the broken phase (or ferromagnetic (FM) phase), where the scalar
field expectation value $v = \langle V_x\rangle$ is 
nonzero, we expect that the boundary fermion for small $y$ gets a 
mass $\propto yv$, but also the gauge boson
acquires a mass $\propto v$. Therefore we cannot decouple the fermion
while keeping the gauge boson light. If we allow the gauge boson to
be massive, it follows from the triviality of the  
Yukawa coupling in this region of the phase diagram, that 
the fermion mass will be of comparable  magnitude.
The remaining option is to choose $\k$ in the symmetric phase (or paramagnetic 
(PM) phase). Here we
expect from experience with the massive Yang-Mills model that the scalar
field can be decoupled from the low energy physics of the fermion-gauge
model,  because deep inside the symmetric phase all scalar excitations will
have masses of the order of the cutoff.
However, since $v=0$, one would also expect the boundary fermion to have
mass zero and therefore not  to decouple, 
which implies that the low energy model would be vectorlike.

An interesting possibility is however that our model might exhibit the strong
Yukawa coupling behavior found in other lattice Higgs-Yukawa models 
[25--27].
It was shown that for strong 
Yukawa couplings in the symmetric phase, such models exhibit another
phase (denoted by PMS) in which the fermion and the scalar field form a 
massive bound state with mass of the order of the cutoff.
We illustrate this  desirable scenario with a possible phase diagram 
at $\k=0$ for our model, for
the Yukawa couplings $y$ and $ye^{-m_0L_s/4}$ of the waveguide boundary and
domain wall fermions respectively, shown in fig.\ 2.

For   $L_s\ra\infty$, the domain wall fermion has  negligible
Yukawa coupling, and we can hope the phase diagram to be similar to that 
of the Yukawa model studied in ref. \cite{SmSwPD}: 
for small $y$ the system starts off
in a weak symmetric phase (PMW); for increasing
$y$ the system comes into a broken phase, because the induced fermion 
interactions are of ferromagnetic nature. 
Then for still larger $y$ the system enters
a strong symmetric phase in which the boundary fermion becomes
massive (denoted by PMS$_1$). 

For finite values of $L_s$ also the
domain wall fermion  gets strongly coupled for large $y$ and this could
take the system into a different symmetric phase (denoted by PMS$_2$),
in which both fermions are massive. Like between the PMW and
PMS$_1$ phases, 
there may be a FM phase separating the PMS$_1$ from the PMS$_2$,
cf. fig.\ 2.
We note in passing that the presence of two Yukawa couplings which are
both proportional to $y$ but differ by a large factor  $e^{m_0L_s/4}$
makes it very difficult to apply a strong coupling expansion in $y$, to
investigate or establish the PMS$_1$ phase analytically.

If the phase diagram of fig.\ 2 would be qualitatively correct
for our model and a PMS$_1$ phase does exist, we could decouple the
unwanted boundary fermion as well as the scalar field:
we can choose the Yukawa coupling $y$ sufficiently strong that the
boundary fermion forms a bound state with the scalar field and acquires a 
mass of the order of the cutoff, whereas the domain wall fermion still is
weakly coupled and remains massless. Sufficiently deep in this PMS$_1$
phase also the scalar field is very massive and should decouple.
When we then turn on a smooth external gauge field inside the waveguide,
the only light particle coupling to it is the right-handed fermion at the
domain wall.
For this scenario to work, we emphasize that the details of the
conjectured phase diagram in fig.\ 2 are not important, 
but only that the PMS$_1$ phase exists.
In the next section we shall investigate the scenario in more detail.

\sect{Mirror fermion representation of the model}
\subsection{Mode expansion}
The colloquial discussion in the previous section can be made more
explicit by rewriting the action as follows. Relabel the right and left 
handed fermion fields, $\Psi_{R,L}^s = P_{R,L}\Psi^s$
as 
\bea
     \psi_{R}^t & = & \Psi_{R}^{s_0-1+t},  \;\; 
     \psi_{L}^t = \Psi_{L}^{s_0-t},  \nn
     \chi_{L}^t & = & \Psi_{L}^{s_0-1+t},  \;\; 
     \chi_{R}^t = \Psi_{R}^{s_0-t},
                         \label{NEWF}
\eea
and the same for  $\Psb_{R,L} = \Psb P_{L,R}$ (note the reversal of $L$
and $R$). The new label $t$ runs from $1$ to $L_t\equiv L_s/2$.
In  fig.\ 1 we 
have indicated this new labeling for the zeromode wave functions shown there.
With our choice 
for $s_0$, $s_0'$ and $L_s$ we can define a domain wall mass for both
fields $\psi$ and $\chi$, which is a step function in $t$ satisfying,
\be
   \mb^t = m^{s_0-1+t} = m^{s_0-t}.
\ee
With this relabeling the two domain wall zeromodes will reside
in the Dirac fermion field $\psi$, whereas the waveguide boundary zeromodes
will reside in $\chi$. After substituting eq. (\ref{NEWF}) into
eq. (\ref{SFUV}) with $U=1$, the action turns into 
\bea
   S_{\j\c} 
       & = &
 \sum_{t=1}^{L_t} \left( \psb^t \ds \psi^t +   \cb^t \ds \c^t +  
     \cb^t( -w + \mb^t)\psi^t + \psb^t( -w + \mb^t)\c^t 
                                                      \right)\nn
  & - & \sum_{t=1}^{L_t-1} \left(
     \psb^t\c^{t+1} + \cb^{t+1}\psi^t \right) +
        \sum_t \left( \cb^t\psi^t + \psb^t\c^t \right) \nn
  & - & y\cb^1(VP_L + V^{\dg}P_R)\c^1
   -   y\psb^{L_t}(V^{\dg}P_L + VP_R)\psi^{L_t}.
                          \label{SMFV}
\eea

In this form, the action resembles that  of an $L_t$-flavor mirror
fermion model in the fashion of ref. \cite{Mont87}, with $\psi$ the
fermion  and $\c$ the mirror fermion field. In fact, for $L_s=2$ 
the hopping terms in $t$ are absent, $\mb^t=0$ and
our model reduces to the mirror fermion model of ref. \cite{Mont87}
with equal Yukawa  couplings for the fermion and the mirror fermion,
and a vanishing single-site mass term.
For $L_s>2$ our model has a more complicated mass matrix (i.e. nondiagonal
couplings among the flavors $s$ or $t$) and if our
model is going to be more successful  in decoupling the mirror fermion 
than the traditional mirror fermion approach, it must come from this
mass term.

The mass matrix for the $L_t$ flavors in our model is not diagonal
but this can be remedied by more rewriting. First we expand the fermion
fields in a plane wave basis, which diagonalizes the Dirac operator and
Wilson term, $\psi^s_x = \sum_p e^{ixp}\pst^s_p$, $\psb^s_x = \sum_p
e^{-ixp}\psbt^s_p$. Here $\sum_p$ is a normalized sum over the 
momenta on the $d$ dimensional lattice, $\sum_p 1 = 1$. Then we can
write,  
\bea
  S_{\j\c} & = &  \sum_{t=1}^{L_t} \sum_p\left( 
i\psbt^t_p \ssl_p \pst_p^t +  i \cbt_p^t \ssl_p \ct_p^t +  
     \cbt_p^t( w_p + \mb^t)\pst_p^t 
  + \psbt_p^t( w_p + \mb^t)\ct_p^t \right)\nn
   & - & \sum_{t=1}^{L_t-1} \left(
     \psbt_p^t\ct_p^{t+1} + \cbt_p^{t+1}\pst_p^t \right) 
  + \sum_t \left( \psbt_p^t\ct_p^t + \cbt_p^t\pst_p^t \right) \nn
  & - & y\sum_{pq}\left( \cbt_p^1(\Vt_{p-q}P_L
        +  \Vt^{\dg}_{q-p}P_R)\ct_q^1 
   +   \psbt_p^{L_t}(\Vt^{\dg}_{q-p}P_L + \Vt_{p-q}P_R)\pst_q^{L_t}
                  \right),
                          \label{SMPFV}
\eea
with $\ssl_p = \sum_\mu \g_\mu \sin(p_\mu)$,  $w_p$ the diagonal form
of the Wilson term, $w_p = \sum_\mu (1 - \cos(p_\mu))$  and $V_p$ the
Fourier transform of $V_x$. For $y=0$ the action has the schematic form 
\be
   S_{\j\c}  =  ( \psbt \; \cbt)\left( \begin{array}{cc}
                           i\ssl   & M^\dg \nn
                            M      & i\ssl  \end{array} \right)
                           \left( \begin{array}{c}
                             \pst \nn
                             \ct  \end{array} \right),
\ee
with $M$ a ($p$ dependent) matrix in flavor space, which can be read
off from eq. (\ref{SMPFV}).
This action can be diagonalized by making unitary transformations 
on $\pst$ and $\ct$,
\beba{lcrlcr}
\o^f & = & F^{\dg}_{ft}\pst^t,\;\; &\ob^f & = &\psbt^tF_{tf},\nn
\x^f & = & G^{\dg}_{ft}\ct^t, \;\; &\xib^f &= & \cbt^tG_{tf},
                      \label{MODE}
\eeea
such that $G^{\dg}_{fs} M_{st} F_{tg} = \m_f\d_{fg}$. 
The matrices $F$ and $G$ are eigenfunctions of $M^\dg M$ and
$MM^\dg$ respectively, labeled by the index $f$:
\be
 (M^{\dg} M)_{st} F_{tf} = |\m_f|^2F_{sf},\Sp   
 (M M^{\dg})_{st} G_{tf} = |\m_f|^2G_{sf}.
\ee
For suitable choices of the phases of the eigenfunctions, the $\m_f$'s 
are  real.  Substituting the mode expansion (\ref{MODE}) into the 
action (\ref{SMPFV}) with the momentum label restored, we arrive at
\bea
  && S_{\j\c} = \sum_{f=1}^{L_t} \sum_p\left( 
   \ob^f_p i\ssl_p \o_p^f +   \xib_p^f i\ssl_p \x_p^f +  
     \xib_p^f \m_p^f\o_p^f + \ob_p^f \m_p^f \x_p^f 
                          \label{SMODE}
                                                      \right)\\ 
  && -  y\sum_{fg,pq}\left[ 
  \xib_p^f G^{p\dg}_{f1}(\Vt_{p-q}P_L + \Vt^{\dg}_{q-p}P_R)G^q_{1g}\x_q^g 
   + \ob_p^f F^{p\dg}_{f L_t}(\Vt^{\dg}_{q-p}P_L + \Vt_{p-q}P_R)
                          F^q_{L_t g}\o_q^g  \right].  \nonumber
\eea

In this representation of the model, it is seen that all fermion modes
$\o^f$ and $\x^f$
interact with the scalar field, but that their effective Yukawa coupling
is determined by the magnitude of their wave function at the waveguide
boundaries $t=1$ and $t=L_t$.
For $y=0$ the model is seen to describe free, degenerate fermions and
mirror fermions with momentum dependent mass $\m^f_p$ (for 
$\m^f_p\not=0$, the eigenstates are $\o^f_p + \x^f_p$ and $\o^f_p -
\x^f_p$). Exactly one flavor, which we denote with $f=0$, has 
$\m^0_p=0$ (up to terms exponentially suppressed in $L_s$)
for $|\ph| < p_c$, where $p_c$ is the critical momentum 
defined in sect. 2.1.  For $r=1$ and $m_0$ close to 1, the
critical momentum is $p_c \approx \sqrt{2}$.
All other $\m^f_p$ and also $\m^0_p$ for
$p$ outside the critical momentum region, are $O(1)$ in lattice units.
This is illustrated in fig.\ 3, where we show the lowest three 
masses as a function of the momentum (again we have chosen $d=2$). 

This shows that for $y=0$  and momenta $|\ph| \aplt p_c$, the model contains 
a massless fermion, $\o^0$, as well as a massless mirror fermion, $\x^0$. 
All other modes ($f\not= 0$) as well as the species doublers have 
a mass of the order of the cutoff. 
The species doublers of the zeromode $f=0$ are massive because 
$\m^0_p$ is $O(1)$ for momenta with $p_\m= \pm \pi$. 
Furthermore,
it is seen in fig.\ 3, that $\m^0_p$ is almost exactly zero 
(it is exponentially small $\propto \exp (-m_0 L_s/4)$) 
until it  quickly rises to nonzero values for $|\ph|>p_c$. 

As was discussed already in sect. 2.1, fig.\ 1 shows the 
$t$-dependence of the zeromodes $F^{t0}$ and $G^{t0}$ 
of the fermion (indicated by $\j$ in the figure) and mirror fermion 
(indicated by $\c$) for  the smallest momenta $|p| =\p/L \ll p_c$. 
It shows that the zeromode for the
fermion is sharply peaked at $t=(L_t+1)/2$, i.e. at the domain
wall and the zeromode for the mirror fermion is localized at the boundary,
at $t=L_t$. The non-zeromodes, which are not shown in this figure, 
are not localized. 

\subsection{Reduced model}

The action (\ref{SMODE}) is an exact representation of the action for
the domain wall fermions. The reason for writing it in this form is
that it reveals, more clearly than the original action, which fermion
modes are important for the low energy physics. 
To shed light on the model for $y\not=0$, we shall exploit this
separation of light and heavy modes 
in order to simplify the model by making a number of approximations,
which we expect to hold for large $L_s$.
First we shall neglect all non-zeromodes. 
This is a reasonable approximation, since these fermion modes
have masses of the order of the cutoff, $\m_p^f=O(1/a), f\not=0$. 
If they would couple strongly
to the zeromodes, they could still be important, but from the Yukawa
interaction in (\ref{SMODE}) one can see that such a coupling involves
the overlap of a zeromode and a non-zeromode at $t=1$ or
$t=L_t$. Since the non-zeromodes are not localized, the value of the
 wave functions $|G_{tf}|$ or $|F_{tf}|$ at any given $t$ is
of order $1/\sqrt{L_s}$ and the 
contribution of $L_s$ internal heavy flavor fermions
is expected to be of order $L_s|G_{1f}|^2/\m_p^f$ = $O(1/\m_p^f)$.

The remaining zeromodes have a momentum dependent Yukawa coupling.
For the fermion $\o^0$ this coupling is proportional to the 
squared absolute value of the wave function
$F_{L_t0}$  which is exponentially small. 
Furthermore, the mixing with the
mirror  fermion is either exponentially small for  momenta $|\ph|<p_c$, 
or the modes are very massive for large momenta $|\ph|>p_c$, 
and we discard such heavy modes in our
approximation. Therefore in this approximation the
model describes a free massless fermion $\o^0$ with momentum cutoff
at $p_c$, and a mirror fermion with Yukawa coupling  to the scalar
field. This Yukawa coupling contains a momentum dependent factor 
$G^{p\dg}_{1 0}G^q_{1 0}$, cf. eq. (\ref{SMODE}). It turns out,
however, that in the momentum range  well below the cutoff $p_c$ this factor
is almost constant  and close to one, and then quickly drops to a small 
value for  $|{\hat p}|\apgt p_c$. This momentum dependence of $|G^p_{1 0}|$ 
is shown in fig.\ 4. 
This justifies the approximation that we also impose the momentum
cutoff on the mirror fermion and neglect the wave function factor in the
Yukawa coupling for $|\ph|<p_c$.

All this leads to a simplified ``reduced'' model, described by the action
\bea
   S^{red} & =  &
  \sum_{|{\hat p}|<p_c}[ i\ob^0_p \ssl_p \o_p^0 + i\xib_p^0 \ssl_p \x_p^0 ]  
  +  y\sum_{|{\hat p}|,|{\hat q}|<p_c} \xib_p^0(\Vt^{\dg}_{q-p}P_R 
+ \Vt_{p-q}P_L)\x_q^0.
                          \label{STOY}
\eea  
Notice that this model differs from the mirror fermion
model of ref. \cite{Mont87} by the absence of a momentum dependent 
mixing term between fermions and mirror fermions and by the presence of the
momentum cutoff $|\ph| < p_c$.

In this approximation, the model shows all the features discussed in 
the previous section. 
In particular we see that the zeromode $\o^0$ which in the full model
is localized at the domain wall,
is nicely decoupled from the mirror fermion $\x^0$
which is localized at the boundary.
The domain wall zeromode has an exponentially small interaction with the 
scalar field,  which we neglected in the action (\ref{STOY}), 
but the mirror fermion couples to the scalar field with strength $y$. 
This mirror fermion will decouple if there exists
a strong symmetric (PMS$_1$) phase  for large $y$ in which the 
mirror fermion and scalar field form a massive bound state.
To summarize, the action (\ref{STOY}) should describe the physics of the
full model for $L_s\ra \infty$, i.e. at the horizontal axis of the
phasediagram in fig.\ 2.

The usual approach to show that such a phase
exists is to write the action in terms of the fermion-scalar
composite field, 
\be
\x_p' = \sum_q(P_R\d_{p,-q} + P_L\Vt_{p-q})\x_q,   \label{OMP}
\ee
which is chosen such that the Yukawa term turns into a mass term for
the  fermion field $\x'$. This fermion does not transform
under the gauge group  $G$, hence we shall call it neutral.
If the momentum cutoff were absent, we could invert this
transformation, $\x_p = \sum_q(P_R\d_{p,-q} + P_L\Vt^\dg_{q-p})\x'_q$,
and a strong coupling approximation of the resulting action for 
$\x'$ would predict a mass  of the order of the cutoff
for this neutral fermion.
However, due to the momentum cutoff such an argument cannot be used for
our model and in fact the transformation (\ref{OMP}) leads to a
nonlocal action for $\x'$. 

The momentum cutoff, which prevents a straightforward  analytic
demonstration that a strong coupling phase exists, makes this model
markedly different from models which are known to have a strong
symmetric phase.  It is somewhat similar, however, to a fermion-Higgs 
model with hypercubical Yukawa coupling \cite{Shig90,LeSh90}.
In these models the Yukawa interaction in
momentum representation contains a momentum dependent factor, coming
from the averaging of the scalar field over the hypercube, which
suppresses the coupling strength for large values of the scalar field 
momentum.  Such models are known not to have a strong symmetric phase.

To summarize this section, we have shown that the domain wall fermion
model can be rewritten as a mirror fermion model, with $L_t=L_s/2$ flavors.
In order to decouple the mirror partner of the domain wall zeromode, 
we must show the existence
of a strong symmetric phase, where the mirror fermion forms a massive
bound state with the scalar field. We have argued that the model  for
large $L_s$ can be simplified to a reduced model with only a fermion and a 
mirror fermion. 
In this model we cannot show the existence of a strong phase using 
standard analytic techniques.
The momentum dependence of the Yukawa interaction (which  gives rise to
the cutoff $p_c$) is more similar to that of
a fermion-Higgs models with a hypercubical Yukawa interaction than to
models with a local Yukawa interaction. Models with a hypercubical
Yukawa interaction that have previously been investigated, are known not
to have a strong symmetric phase.

Of course the similarity to hypercubically coupled fermion-Higgs models
does not prove that a strong phase is absent in our model, and we shall
search for it with numerical methods. The most direct way to show the
existence of a strong coupling phase, is by measuring the mass of the 
boundary (mirror) fermion for strong Yukawa coupling.
In the next section we shall study the fermion masses, 
both for the domain wall fermion and the boundary fermion, 
in the quenched approximation. 
We shall compare these results with the masses found from the reduced model.

\sect{Fermion spectrum: numerical results}

In order to substantiate the discussion in the previous sections, we shall
compute the fermion spectrum of the full domain wall model in the
quenched approximation. We expect from experience gained with
other fermion-Higgs models, that the presence of a strong coupling
symmetric phase   if it exists, can already be shown within the 
quenched model. 
In the quenched approximation there are no real phase transitions
separating the PMW, PMS$_1$ and PMS$_2$ phases of fig.\ 2.
One expects, however, that the FM phase separating these phases in the
unquenched model, now turns into a cross-over region, which
separates regions of the quenched phase diagram with different (weak
and strong coupling) behavior. In the following we shall refer to
these regions as weak and strong coupling phases, as in sect. (2.3).

For weak Yukawa coupling the  mirror fermion mass in the quenched
approximation is expected to behave as $m_F\approx y v$,
where the scalar field expectation value $v$ is zero in the
symmetric phase and nonzero in the broken phase. A strong coupling
symmetric phase would lead to $m_F = yc(\k)$, with $c(\k)$ a function
of the scalar field hopping parameter $\k$. Typically $c(\k)>v(\k)$, it
decreases with $\k$, and
in particular it is nonzero and $O(1)$ in the symmetric phase.
For instance, in the model of ref. \cite{SmSw80}, $c(\k) \approx 1/z(\k)$ 
with $z^2 \propto {\rm tr}\langle V_xV^\dg_{x+\m}\rangle$. 
For the domain wall fermion mass we expect $m_F\approx 0$ for all $\k$ 
 and $y e^{-m_0L_s/4}\ll 1$.

For this numerical study we use the domain wall model in $2+1$
dimensions with  gauge group $G$=$U(1)$, but we keep the gauge 
fields in the global symmetry limit $U=1$. 
The scalar field action (\ref{SV}) then is that of an
XY model, and in the quenched approximation, where  the scalar field
dynamics is determined solely by the action (\ref{SV}), there is a vortex
phase and a spinwave phase.
The Kosterlitz-Thouless phase transition is at $\k=\k_c\approx 0.5$ 
in our convention for the action.
Of course spontaneous symmetry breaking does not really occur in this
two dimensional model, but on a finite lattice the field expectation
value $v$ shows a behavior similar to that in a model with spontaneous
symmetry breaking: it is nonzero and $O(1)$ for $\k>\k_c$ and then
quickly drops to a small (nonzero) value for $\k<\k_c$. For increasing
volumes $v$ becomes closer to zero for $\k<\k_c$ but also in the
spinwave phase it decreases slowly, such that in the limit of infinite
volume $v=0$ everywhere, as it should. 
We emphasize that in not too large volumes, in which there is a clear
distinction between the value of $v$ in the vortex and spinwave
phases, we expect a similar relation between fermion mass and $v$ as
in a four dimensional model with spontaneous symmetry breaking.
Hence we shall refer to the vortex phase as the symmetric phase
and to the spinwave phase as the broken phase.

To find the fermion masses, we have measured the propagator in
momentum space,
\be
  S^{st}(p) = L^{-2}\sum_{xy} e^{ip(x-y)}\av{\Psi^s_x\Psb^t_y},
                              \label{PROP}
\ee
with $L^2$ the two dimensional lattice volume. Optimally, one should
measure the full matrix $S^{st}(p)$ in flavor space, for a number of
small momenta $p$ and from that compute the 
massive and massless eigenstates. 
However, the number of flavors typically is large  (we use for instance
$L_s=26$), and it is impractical to compute the propagator matrix for
all flavors $s$. Since we are only interested in the masses of the 
light states and since we know that these states are localized either 
at the domain wall or at the waveguide boundary, it is sufficient to compute 
only $S^{ss}$ for selected $s$-values $s=1,s_0-1,s_0,L_s/2+1,s_0'$ and 
$s_0'+1$. In fact, we know from the discussion in the previous section
that the mirror fermion is localized near
$s=s_0$ and we need only consider $s=s_0-1$ and $s_0$ if we are
interested only in the mirror fermion mass. 

The parity symmetry of eqs. (\ref{PARITY}) and (\ref{VPARITY})
can be used, after averaging over the scalar field, to
relate certain $RR$ and $LL$ components of the fermion propagator:
\be
S_{RR}^{st}(p)=S_{LL}^{L_t+2-s,L_t+2-t}(Pp),
			\label{RRLL}
\ee
where $Pp$ is the parity reflected two-momentum.
We have used this relation to average over the appropriate 
$RR$ and $LL$ components, in order to increase statistics.
For the $RR$ or $LL$ component of a free fermion propagator we expect 
\be
S(p)_{RR(LL)}=    
-iZ_F(\sin(p_1) -(+)i\sin(p_2))/(\sum_\mu \sin^2(p_\m) + m_F^2),
               \label{PROPS}
\ee
where $Z_F$ is a wave function renormalization constant and $m_F$ is
the mass. In fig.\ 5a we have plotted the inverse of the 
averaged $RR$ and $LL$ components,
as a function of $\sum_\m \sin^2(p_\m)$. 
We used a lattice of size $L^2L_s=12^2 26$ with Yukawa coupling
$y=0.5$, at $\k =0.5$ near the phase transition and $m_0=1.1$. 
We used antiperiodic boundary conditions for the fermions in the 
$t$-direction.  The data have been normalized such that the
slope (determined from the first and second point) is one. 
The straight lines are  $\c^2$ fits to the data and the good quality of
these fits shows that the fermions are (nearly) free.

To see if the 
reduced model resembles the full model also in a quantitative
way, we have computed the inverse mirror fermion 
propagator in  this model. This result is shown in fig.\ 5b, 
again normalized to slope one.  The normalized mirror fermion propagator
is in good agreement with  the one computed in the full model.

As anticipated the $RR$ component of the domain wall fermion at
$s=L_s/2+1$ (and the $LL$ component at $s=1$) has zero mass.
The mirror fermion modes at $s=s_0-1$ (the $RR$ component) and
at $s_0$ (the $LL$ component) have a small mass which is
consistent with $m_F = yv$. All other components are seen to have
a mass of order one in lattice units. In the same fashion we have
computed the fermion masses at other values of $\k$ and $y$.
In all cases we found that the domain wall fermion remains massless.

Of course the most interesting results are those for the mirror fermion
mass at small $\k$ in
the symmetric phase and at large values of $y$. Unfortunately, the
data here are subject to large statistical fluctuations. Even after
averaging over 3000 scalar field configurations at $\k=0.1$ and $y=10$ 
we found that the errorbars on the propagator are comparable with the 
signal. The reason is that the propagator itself is very small, which
prohibits a reliable analysis of the fermion propagator for such values 
of the couplings.

Instead of using $\k\approx 0$ such that $v\approx 0$, we can also
look for strong coupling behavior at larger $\k$. In the broken phase
the weak and strong regions are less pronounced, but
the presence of a nearby  PMS phase should still show 
up in a deviation from the relation $m_F\approx yv$. Since $v$ decreases for
$\k\searrow \k_c$, a characteristic feature of weak coupling behavior is a
fermion mass which decreases as $\k\searrow \k_c$. As mentioned above,
strong coupling
behavior would show up through an opposite trend of the fermion mass
as a function of $\k$, increasing towards the phase transition.
In fig.~6 we show the $\k$ dependence of the waveguide fermion
mass at fixed $y=2$. One recognizes the typical weak coupling behavior
of the mass. From experience with other models
we expect that $y=2$ is already a strong coupling. 
For larger $y$ it is difficult to measure the fermion mass reliably, 
because it is comparable to the cutoff in the range $\k\apgt 0.5$. 
In fig.\ 7 we have plotted
the $y$ dependence of the mass for fixed $\k=0.5$. 
Strong coupling behavior should show up as a relative increase of the mass 
compared to the weak coupling trend. From fig.\ 7, however, we can 
at most infer a relative decrease of the mass for $y>1$.
For comparison, we  have also plotted the line $yv(\k=0.5)$ in this figure.

The crosses in figs.\ 6 and 7 are the masses
obtained from the  reduced model. One sees the same qualitative
behavior as in the full model, but the masses are systematically higher
(except when $m_F\apgt 2$, which is beyond the cutoff, 
where, in the full model, mixing with all the other heavy modes
presumably becomes important). 
This difference  may be due to the momentum dependence of the
fermion wave function in the full model. 
For increasing momentum and masses closer to the cutoff, we expect
the wave function to spread out and the overlap at the waveguide boundary
to decrease. This implies that the residue $Z_F$ of the fermion propagator
(\ref{PROPS}) is not constant but decreases with increasing momentum.
This also leads to an underestimate of the fermion mass in the full model.
Keeping these systematic effects in mind, we consider the results of the 
reduced model in  reasonably good agreement with the full model.

Even though at this stage we do not yet find a conclusive answer for $\k$ 
deep in the symmetric phase and large Yukawa coupling, the results shown 
in figs.~5, 6 and 7, are
consistent with the  weak coupling mass relation $m_F=yv$. Also the
awkward behavior of the model at small $\k$ and large $y$ is not
what we expect from a model in the strong coupling phase. Only
in the transition region between the two regimes we expect large
statistical fluctuations, but after the bound state has formed, the
model should describe weakly coupled massive Dirac fermions, whose mass 
should be easy to measure.  

\sect{Search for a strong coupling symmetric phase}
\subsection{Eigenvalue spectra}

The results for the mirror fermion mass described above are
very suggestive but did not give a
conclusive answer to the question whether a strong coupling phase exists in
our model. Therefore we will attempt to approach this problem from
a different angle in this section.  
The idea here will be that 
the presence of a strong coupling phase shows up in the distribution
of the eigenvalues of the fermion matrix \cite{SSev}. 

Of course, we would like to look at the eigenvalues of the domain wall 
fermion matrix directly. This $2L^2L_s\times 2L^2L_s$ 
(the factor $2$ comes from the Dirac index) matrix $M$ is obtained by 
writing the action (\ref{SFUV}), with $U=1$, in the form 
$S = \psib M \psi$. 
However, it is unpractical to study $M$ directly, because this 
(nonhermitian) matrix is too large to handle numerically on reasonably sized
lattices, and it is not clear
what to expect for the distribution of the eigenvalues for $M$
in the representation following from eq. (\ref{SFUV}). 
Only in the representation that diagonalizes the mass matrix in flavor 
space we should expect similarities with the eigenvalue spectra of 
free fermions (for small $y$) with momentum dependent masses.

This suggests that we use the reduced model, which is formulated
in terms of these mass eigenstates, and which contains much less
degrees of freedom. The reasonable agreement of the results for
the fermion masses discussed in the previous section supports this
strategy. In the reduced model we can compute the distribution of
the eigenvalues of the fermion matrix at small, intermediate and large
Yukawa coupling. Then we can compare these eigenvalue spectra with those
obtained in models for which a strong coupling phase is known to
exist or to be absent. As such reference models we use a model with
naive fermions with local Yukawa coupling, which has a strong coupling phase,
and the same model with hypercubical Yukawa coupling, which has no strong
coupling phase. The actions for these models, which we shall refer to
as the $Y_{lc}$ and $Y_{hc}$ models are,
\bea
  S_{lc} & = & \sum_{xy}\psb_x \ds_{xy} \psi_y 
                 + y\sum_x\psb_x(V_x P_R + V^*_x P_L) \psi_x ,
 \\ 
  S_{hc} & = & \sum_{xy}\psb_x \ds_{xy} \psi_y 
  + y\sum_{x}\quart\sum_b\psb_x(V_{x-b} P_R + V^*_{x-b} P_L) \psi_x.
                      \label{YMODEL}
\eea
The sum over $b$ in the hypercubical Yukawa interaction runs
over the four corners of the elementary plaquette, $b_\m = 0,1$.
After Fourier transforming the $Y_{hc}$ model, we find 
\be
  S_{hc} = \sum_p \psbt_p i\ssl_p\pst_p + y\sum_{pq} h_{p-q}
           \psbt_p(\Vt_{p-q}P_R + \Vt^*_{q-p}P_L)\pst_q
\ee
which contains a factor $h_{p-q} = \prod_\m e^{i(p_\m-q_\m)/2}
\cos((p_\m-q_\m)/2)$, 
which goes to zero for large momenta $|p_\m-q_\m|\ra \pi$ of the $V$ field.

The results of this comparison are presented in fig.\ 8.
Fig. 8a contains the spectra of our reduced domain wall
model at $\k=0.1$ and $y=0.2$, $1.0$ and $4.0$.  We have plotted the
eigenvalues obtained from $5$ quenched scalar field configurations, with
lattice size $L=12$.
The figure shows that the scattering of the eigenvalues 
caused by the strongly fluctuating scalar field increases with 
increasing $y$, as expected.
However, there is no sign of a qualitative change for larger $y$.
Also for Yukawa couplings $y>4$  we found that the spectra do not change
qualitatively, they just scale proportionally to $y$.

This can be contrasted to the $y$ dependence of the eigenvalue spectra
in the $Y_{lc}$ model shown in fig.\ 8b. Here we see an
increase of the fluctuations for $y$ growing from $0$ to $1$, then
the eigenvalues $\l$ start to rearrange 
themselves  along the boundary of a crude circle, which cuts the real
axis at approximately $\pm y$, such that the region
around the origin becomes depleted of eigenvalues. This signals the
existence of a strong coupling phase for $y\apgt 1$, cf. ref. \cite{SSev}. 

The spectra of our reduced model do not show such a qualitative change for
large $y$ and are much similar to those shown in fig.\ 8c,
which were obtained from the $Y_{hc}$ model, and which does not have
a strong coupling phase.  Since we expect the reduced model to be 
qualitatively similar to the full  model for large $L_s$, this result
casts serious doubt on the existence of a strong coupling phase in our
domain wall fermion model.

The properties of the eigenvalue distribution of the fermion matrix are
also reflected in
the behavior of the conjugate gradient (CG) inversion. For small
Yukawa coupling and using anti-periodic boundary conditions to regulate
the zeromode for the fermions, we expect a rapid convergence of the
CG inversion on our relatively small lattice. Then for
increasing $y$  the inversion rate should deteriorate,
i.e.\ the number of CG iterations to reach the solution
to a given precision will increase. 
If there is a strong coupling phase, the number of iterations
reaches a maximum at the cross-over to the strong phase and then
decreases again, because in the strong coupling phase the composite
fermions are again weakly coupled and massive. 
In ref. \cite{SmSwPD} it was found
that the number of CG iterations provided an accurate
indicator for the location of the cross-over and the existence of the
strong coupling phase. 

In fig.\ 9 we have plotted the number of
CG iterations as a function of $y$ at $\k = 0.1$ ($L=12$, $L_s=26$).
One recognizes the expected rise of the number of iterations when
$y$ increases from $0$ to $\approx 1.5$. But, unlike what one expects for a
model with a strong coupling phase, there is no decrease for large
$y$. Also after $y=2$ the number of iterations keeps rising, albeit
at a slower rate and with larger fluctuations than at small $y$.
For comparison we have also plotted the $y$ dependence of the
number of CG iterations obtained in the $Y_{lc}$ and $Y_{hc}$ model.
In the $Y_{lc}$ model, which has a strong coupling phase, the number of CG
iterations clearly shows a peak at $y\approx 1$; in the $Y_{hc}$ model,
which has no strong coupling region, we see a behavior
similar to that of our domain wall model. 

We do not have an analytic method to establish the existence or
nonexistence of a strong coupling phase in our model, but by comparing
the $Y_{lc}$ and $Y_{hc}$ models, one might conjecture that a strong
phase can only exist if the fermion and scalar modes are coupled strongly
over the full momentum range, including the high momenta modes with
$p_\m$ near $\pm \pi$. This is the case in the $Y_{lc}$ model, but both 
in the $Y_{hc}$ model and in our domain wall 
model the Yukawa coupling is suppressed for large momenta, though the
details of the momentum dependence are different in the two models.

\subsection{Dynamical fermions}

We have also attempted an unquenched simulation of the model. 
A direct simulation of the model with the action (\ref{SFUV}),
$S = \psb M \psi$, is not feasible, because $\Det M$ is not
positive definite. Therefore we have simulated instead a model with
an extra fermion field $\c$ added, with action $S=\cb M^\dg \c$, such
that the fermion determinant is given by $\Det(M^\dg M)$.  
In this model we can use a hybrid Monte Carlo algorithm to include 
the fermions, because the fermion determinant is now manifestly 
positive definite.
  
To look for a strong coupling symmetric phase, one
should use a small value of $\k$, such that the
model is in the symmetric phase at $y=0$. For $y>0$ we expect that
the system switches to the broken phase. This disappearance of the
symmetric phase for arbitrarily small but nonzero $y$ is a special
feature of two dimensional Yukawa models, cf. e.g. \cite{DeFo93}
and  references therein.  Then for larger $y$ we expect to find 
a symmetric phase if such a phase exists.
Unfortunately, the system turns out to be extremely hard to simulate
numerically for small $\k$.
For $\k=0$ and small $y$ we could still measure nonzero field
expectation values, but for increasing $y$ we had to
decrease the trajectory length progressively more, 
to unacceptably small values 
(e.g. at $y=1$ we had to use a step size $dt=0.01$ with $10$ steps per 
trajectory to maintain an acceptance rate larger than 75\%, and 
at $y=10$ we had to use $dt = 0.001$). 
This results in huge autocorrelation and equilibrium times, which make 
a realistic simulation unfeasible for values of $y$ 
in the interesting region. 
Presumably this is due to large fluctuations in the eigenvalues of
$M^\dg M$, as is suggested by the spectra of the reduced model shown in
fig.\ 8. In fig.\ 9 we have also plotted the number of CG iterations
required for inversions in the unquenched model at $\k=0$
(full triangles). This shows
the same steady increase with $y$ as found in the quenched model.
Also a  tentative run at $y=10$, showed none of the
improvement we would expect after moving into a strong coupling phase.

\sect{Other ways to couple the gauge fields}

The model we have studied in this paper 
resulted from an attempt to couple gauge
fields to the chiral mode on only one of the domain walls, while
preserving gauge invariance. This gave rise to an extra scalar field
and additional mirror fermion modes at the waveguide boundary.
These mirror fermion modes do not seem to decouple, in other words,
there does not seem to exist a region in the phase diagram where the
mirror fermions have masses of the order of the cutoff, while the 
zeromodes at the domain walls remain light.
However, one might wonder whether the gauge field cannot be coupled to
the fermions  in a different way, which  avoids these complications. 
Unfortunately, it 
appears to be difficult to find a different approach without obvious
flaws.  Let us briefly discuss the original proposal \cite{Kapl92},
and its relation to the model studied in this paper.

In the original proposal gauge fields were put on all
links of the $d+1$ dimensional lattice, i.e. the gauge field was taken
to be a full, $d+1$-dimensional gauge field. The action for these gauge
fields was chosen as
\be
   S(U) = \sum_{x,s}\left( \b \sum_{\m,\n=1}^d\RE 
      \Tr (U^s_{\m x}U^s_{\n x+\mh}U^{s\dg}_{\m x+\nh}U^{s\dg}_{\n x})
 + \b_{d+1}\sum_\m \RE \Tr (U^s_{\m x}V^s_{x+\mh}U^{s+1\dg}_{\m x}V^{s\dg}_x)
             \right).
			\label{GAUGE}
\ee
The gauge field in the $d+1$  direction is denoted by $V$.
By choosing the coupling  $\b_{d+1}$ for the extra field $V$ 
sufficiently
different from the plaquette coupling $\b$ for the gauge fields $U^s$, 
it was hoped that at the domain wall the gauge field dynamics would
still be $d$-dimensional at scales much below the cutoff.

Eq. (\ref{GAUGE}) can be viewed as  the action for 
a number of $d$-dimensional gauge
fields $U^s$ (labeled by $s$), coupled to equally many unitary scalar 
fields $V^s$.  For $U^s=1$ we have just $L_s$ independent 
nonlinear sigma models in $d$ dimensions, each with a critical point at
$\b_{d+1}=\b_c$.  For each $s$, the global symmetry group is $G\times
G$, with $V^s$ transforming as
\be 
V^s \ra g^s V^s(g^{s+1})^\dg,              \label{GLOBVS}
\ee
with the $g^s$ in $G$.  The full symmetry group 
${\cal G}=G^{L_s}$
is gauged by the $d$-dimensional gauge fields $U^s$.  The hopping terms
in the $s$-direction in the fermionic part of the action look like the
Yukawa terms in our model, eq. (\ref{SFUV}) (with $y=1$):
\be
  S_{\J V} = - \sum_s 
            (\Psb^{s}V^sP_L\Psi^{s+1}+\Psb^{s+1}V^{s\dg}P_R\Psi^{s}).
                     \label{SYUK}
\ee

For $U^s=1$ and $\b_{d+1}<\b_c$, the symmetry $\cal G$ is unbroken, and 
$\av{V_x^s}=0$ for all $s$.  
It is then easy to see that in a mean field approximation,
where $V^s_x$ is replaced by $v=0$, the fermion action is that of $2$
massless Wilson fermions and $L_s-2$ Wilson fermions with mass 
$\approx m_0$, which
are decoupled from each other, and vectorlike in $d$ dimensions.
This has been investigated in more detail in ref. \cite{Kort93}.

If we now take $\b_{d+1}>\b_c$, the group $\cal G$ breaks down to its 
diagonal subgroup, $G^{L_s}\ra G$, and only one gauge field remains
massless.  The other gauge fields would get a mass $\propto v=
\langle V\rangle$, and could be made very massive by choosing $v$ at the
cutoff.  The fermion hopping terms in the $s$ direction would survive
(in mean field), and we would find the usual zeromodes at both domain
walls.
However, the massless gauge field is independent of $s$ and couples equally
to the modes at the domain and anti-domain walls, again rendering the model
vectorlike.  

In a sense then, the model which we studied in this paper, is an 
improvement on this situation. 
Formally, our model corresponds to choosing $\b_{d+1}=\infty$ for all
but two $s$-slices (where we set $\b_{d+1}=\k$), forcing the 
$d+1$-dimensional gauge field to be $d$-dimensional.  
There is no  interaction with the antidomain wall if we choose
the plaquette coupling $\b = \infty$ outside the waveguide.
 
In our model we have chosen an $s$-dependence of $\b_{d+1}$ and $\b$,
(and of the fermion hopping parameter in the $s$-direction by the
introduction of a Yukawa coupling $y\not=1$ at the waveguide boundary),  
which we think had the best chance of producing a chiral model.
Of course, a more
general $s$-dependence is possible, but we do not believe that this
will improve the situation as described in this paper.

\sect{Summary and conclusion}

In this paper we have considered a gauge theory with 
domain wall fermions in a 
finite volume with a right-handed zeromode living at the domain wall and a
left-handed zeromode at the anti-domain wall.  
The right-handed mode at the domain wall is coupled
to a four dimensional gauge field which
is confined to a waveguide around this domain wall.
The left-handed mode at the anti-domain wall remains uncoupled \cite{Kapl93}.
The fermion hopping terms across the waveguide boundaries break 
gauge invariance, which is restored by promoting these hopping terms 
to Yukawa couplings in a way similar to the way fermion mass terms are 
made gauge invariant in the Standard Model.  
This leads to the introduction of a scalar field which lives only at the
boundaries of the waveguide. There are two parameters in this model 
associated with 
this scalar field, a Yukawa coupling $y$ and a hopping parameter 
$\kappa$ (or equivalently a mass) for the scalar field.  
In our numerical work we have studied the scalar-fermion dynamics 
in the model with U(1) gauge symmetry in $2+1$ dimensions. The gauge
fields, which can be treated perturbatively, are switched off and 
we mainly used the quenched approximation.

For vanishing Yukawa coupling the regions inside and outside the waveguide 
decouple from each other and from the scalar field. Therefore
one would expect that new chiral zeromodes show up at the waveguide 
boundaries.  This is indeed what happens: there is a left-handed mirror mode 
just on the inside of one of the waveguide boundaries, 
and a right-handed mirror mode on the outside (cf. fig.\ 1).
The inside mirror fermion couples to the gauge field in the waveguide, 
resulting in a vectorlike theory. 
To show explicitly that our  domain wall fermion model can be interpreted as
a mirror fermion model, one can view the extra dimension as a flavor
space. The hopping and single site terms in the extra dimension then
generate a mass matrix, which is not diagonal in flavor space. 
By diagonalizing this mass matrix
at $y=0$, one recovers the massless domain wall modes as well as the
massless mirror partners at the waveguide boundary. All other modes have
masses of the order of the cutoff. For $y\not=0$ all modes have
Yukawa interactions with the scalar field, 
proportional to $y$ and to the magnitude of the wave function of the
particular mode at the waveguide boundary.
Since the wave function of the domain wall mode is exponentially
small at the waveguide boundary, its Yukawa interaction is very weak,
even at large values of $y$;
the mirror mode, however, interacts strongly with the scalar field.

The crucial question is then whether the mirror fermion can be decoupled.
A favorable possibility would be that for large 
Yukawa coupling, the mirror fermion at the waveguide boundary forms a bound
state with the scalar field, with a Dirac mass of the order of the cutoff, 
while the gauge symmetry remains unbroken.  
Such a strong coupling behavior is known to exist in many fermion-scalar 
models.  
In particular a strong symmetric or paramagnetic (PMS) phase has been
established in these models.
The key point in our model is that only 
the mirror fermion should become heavy, while the modes at the domain walls 
should remain massless.  
This would be conceivable, because the mirror mode couples
much more strongly to the scalar field than the domain wall mode.

One might ask whether such a scenario is excluded by the simple 
consideration that the massless mirror mode is required to cancel the
anomaly generated by the domain wall mode.  
We think that this is not the case.  
If one turns on a smooth external gauge field,
a Goldstone-Wilczek current will carry charge away from the 
domain wall.
However, since no gauge field is present outside
the waveguide, this current vanishes in that region, and the charge
will have to be deposited somehow at the waveguide wall.  Of course, if 
massless mirror fermions are present, they will do the job, much as the 
antidomain wall 
zeromodes did in the case without a waveguide, but rather with an external
gauge field present throughout space-time \cite{Kapl92,Jans92,GoJa92}.  
However, an alternative possibility is that a Wess-Zumino current 
carries the charge at the waveguide without any massless fermion
modes being present.  In the $V=1$ gauge (where $V$ is the scalar field), 
a charge density of the form $j_0\propto \epsilon_{0ij}\partial_iA_j$ can
be nonzero due to the discontinuity of the gauge potential $A$ at the 
waveguide boundary.

In our numerical work we have used several approaches to search for the
existence of a PMS phase, but never found an indication that it exists.
Our best evidence that it is absent, comes from computations
of the mirror fermion mass at values of $\kappa$ near the phase
transition to the symmetric 
phase, where the dependence of the mirror fermion mass on $\k$  and $y$
is consistent with the weak coupling mass relation $m_F=yv$ even for
large values of $y$ ($v$ is the scalar vacuum expectation value). 
No sign of the behavior
typical of a strong Yukawa coupling region was found (cf. figs.\ 6 and 7).
For all values of the Yukawa coupling
that we have considered, the domain wall zeromodes 
remain massless and unaffected by the Yukawa interactions.  
It would of course be nice to 
directly measure fermion masses deep in the symmetric phase,  for
small $\kappa$ and large $y$. 
However, in this parameter region the signal for the propagator 
disappears in the noise and we have not been able to
obtain data with small enough errors to draw any definite conclusion 
about the mirror fermion mass.

Since direct computations of the mirror fermion mass run into numerical
difficulties, 
we have also studied a ``reduced'' model, which contains only the mirror
fermion interacting with the scalar field. It is
obtained from the full model for large $L_s$ (the extent  in the extra
dimension) by 
discarding all fermion modes with masses of the order of the cutoff.
Since the Yukawa coupling of the domain wall zeromode is exponentially
suppressed in $L_s$, this mode can also be discarded.
The momentum dependence of the mirror fermion wave function
is such that the effective Yukawa coupling for this mode is suppressed
for momenta larger than the critical momentum $p_c$ and hence we also
discard these large momentum modes. We computed the eigenvalue spectrum of 
the fermion matrix for this reduced model as a function of $y$, and 
compared this with typical eigenvalue spectra for simple
Yukawa models.  The eigenvalue spectra of such models are very different 
for large $y$, depending on whether a PMS phase does or does not exist 
\cite{SSev}.  The eigenvalue spectrum of the reduced model shows no sign 
of a PMS phase (cf. fig.\ 8).

Of course, one would like to study the eigenvalue spectrum of the full model
directly.  This was not possible due to the prohibitive amount of 
computer resources that would be needed.  
We believe however, that the ``reduced'' model captures
the essential features of the full model, one of which is the existence of an
effective momentum cutoff at $p_c$. 
This belief is supported by a reasonable agreement
between light fermion masses computed in the full and reduced models.

It appears that the existence of an effective momentum cutoff $p_c$ in the 
theory, is the underlying reason for the
failure to find a PMS phase.
In the domain wall approach the fermion doublers are
decoupled by making them heavy which implies that these modes are
not bound to the domain wall or waveguide boundary, as is the case with the
light modes. This implies that for large momenta near the doubler
momenta $p_\m = \p$, the wave function of the boundary mode
will be spread out in the extra dimension and it
will be small at the location of the scalar field. Therefore the effective
Yukawa coupling for these large momenta modes is necessarily small. However,
this suppression of the Yukawa coupling for large momenta then
prevents the formation of fermion-scalar field bound states necessary to
have a strong coupling phase.  As a result, the mirror fermion at the 
waveguide boundary stays light, and renders the theory vectorlike in the
scaling region.
If this picture is right, it points at a fundamental problem for
domain wall fermions with a waveguide, not just for the two dimensional 
quenched U(1) model investigated here. 

The results discussed above were obtained in the quenched approximation. The 
unquenched model, assuming the decoupling of the boundary fermion would have 
been
successful, would describe a single right handed fermion
interacting with a U(1) gauge field. This model is anomalous and one
might fear that our unfavorable results are a reflection thereof.
This, however, is not likely, because we can also think of our model
as the quenched approximation of a vectorlike model, obtained by
adding an extra mirror fermion: writing the original action (\ref{SFUV})
as $S = \psb M \psi$, we can add an extra fermion field $\chi$ with
action $S = \cb M^\dg \chi$. For the additional $\chi$ fermion, the
handedness of the zeromodes at the domain wall and waveguide boundary is
reversed and the model is now anomaly free. In the
quenched approximation, however,  the extra fermion is irrelevant and
this model reduces to the one studied here.

We have performed some unquenched simulations in the model with the
extra fermion included, using a hybrid Monte Carlo algorithm.  The results
are inconclusive due to the very large autocorrelation and equilibration
times, but do not contradict the conclusions described above.

All our numerical computations have been carried out within a restricted 
range of Yukawa couplings, as typically the signal to noise ratio
deteriorated prohibitively for large values of $y$ in the
symmetric phase.  Therefore, it is not 
logically excluded that some PMS like behavior might be found at values of
$y$ beyond $y\approx 10$ or so.  In particular, we have not tried to 
investigate the existence of a PMS$_2$ phase as described in section 2.3.
This phase would not be interesting, however, for the construction of a 
chiral gauge theory, since also the domain wall zeromodes would be 
strongly coupled to the scalar field and neutral with respect to the gauge 
charge.

To summarize, we believe that all the evidence presented in this paper
--- the close resemblance to a mirror fermion model with hypercubical
Yukawa interaction, 
the $\k$ and $y$ dependence of the mirror fermion mass where we could
measure it, the distribution of the eigenvalues  in the reduced model 
for large $y$ and the behavior of the conjugate gradient algorithm ---
indicates that a PMS phase does not exist in our model and 
that the mirror fermions, which exist as a consequence of the 
introduction of a waveguide, cannot be made heavy. 
Therefore, a vectorlike gauge theory will
result when gauge interactions are turned on.  
In view of the discussion in section 6, we expect that  
this negative result is quite general for domain wall fermion models
in which the volume in the extra dimension is kept finite  
at any stage in the definition of the model.

\subsubsection*{Acknowledgements:}
We would like to thank D. Kaplan for many stimulating discussions and
valuable suggestions.
We would also like to thank G. Bodwin, A. De, A. Gonzalez-Arroyo, 
J. Kuti and R. Narayanan for discussions.  
M.G. would like to thank the Physics Department of UC San Diego, and K.J. 
that of Washington University for hospitality.
This work is partially supported by the DOE under grants DE-FG03-91ER40546
and  DOE-2FG02-91ER40628, and by the TNLRC under grant RGFY93-206.
The numerical computations were performed on the Cray Y-MP8/864  at the
San Diego Supercomputer Center.
%
%
%

\newpage
\subsection*{Figure captions}
\begin{description}
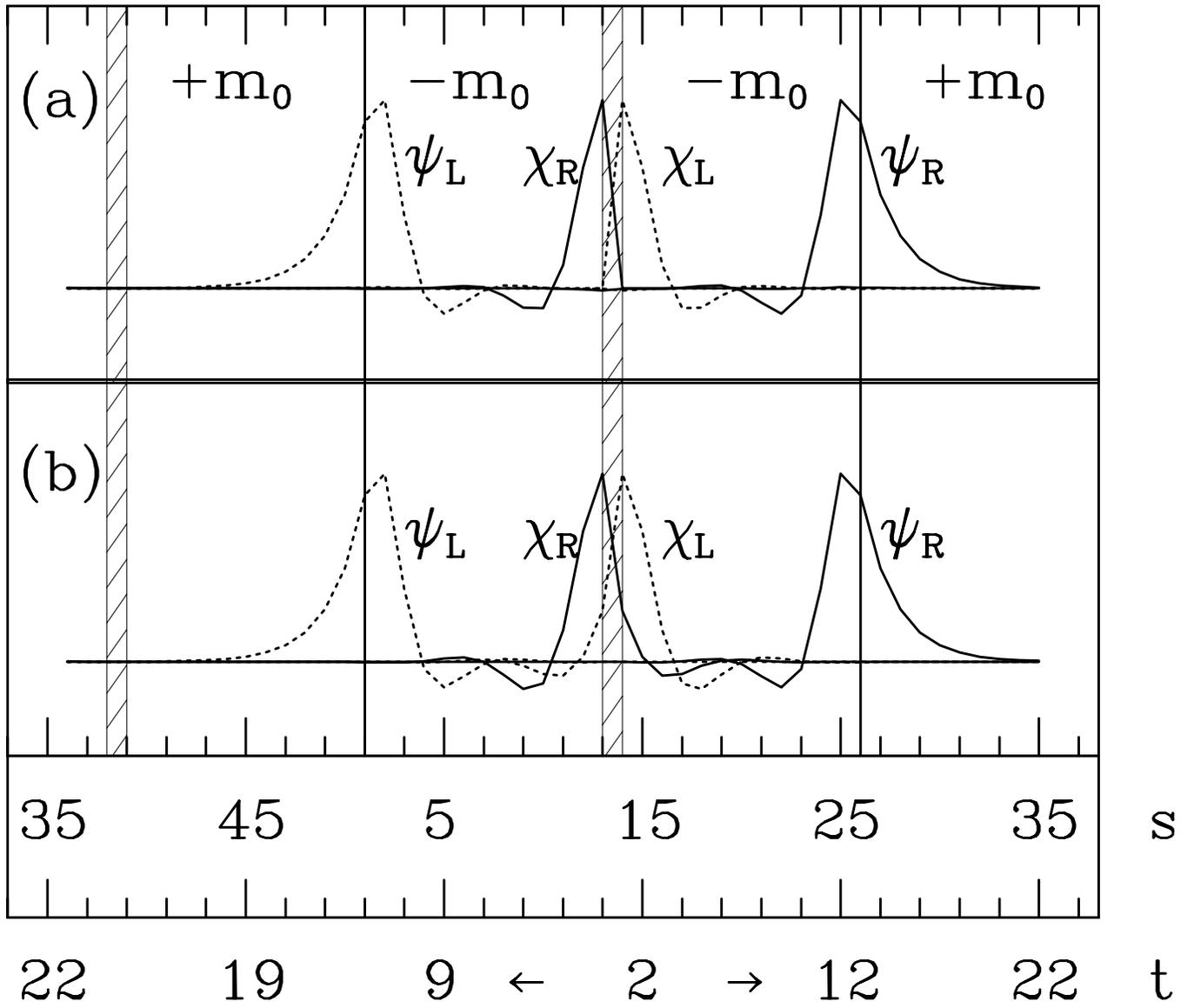
 \item[Fig. 1:] Wave functions of the four lightest modes with
                momentum $|p|=\p/L$, on a $2+1$ dimensional lattice with 
                $L=18$, $L_s=50$ and $m_0=1.1$. The
                solid (dotted) lines represent right (left) handed
                components. The $\psi$ and $\chi$ are the fermion and
                mirror fermion located at the domain wall (vertical bar)
                and waveguide
                boundary (dashed band) respectively. We indicate both
                the $s$ and $t$ labeling defined in eq. \protect{(\ref{NEWF})}.
                Figure a is for $y=0$, figure b for y=0.5.

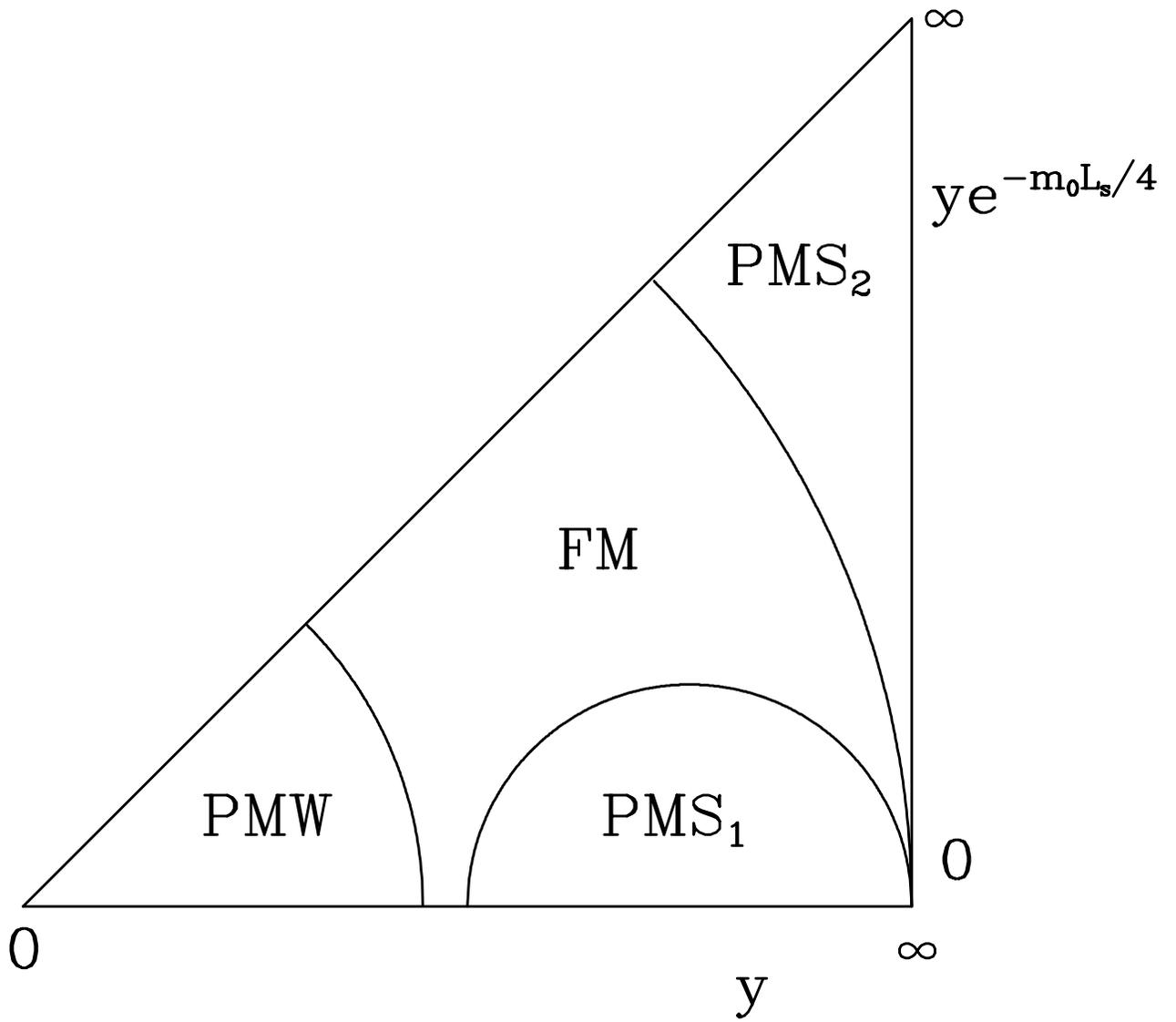
 \item[Fig. 2:] Sketch of a phase diagram which would make the the domain 
                wall fermion model successful. The $y$ and $ye^{-m_0L_s/4}$
                indicate the effective Yukawa couplings for the domain wall
                and waveguide fermion respectively and $\k=0$. The various 
                phases are explained in sect. 2.3.

 \item[Fig. 3:] Momentum dependence of the three lowest mass eigenvalues
                $\m^f$ at $y=0$, obtained on a $2+1$ dimensional lattice with
                $L=16$, $L_s=50$ and $m_0=1.1$. The domain wall fermion
                and waveguide fermion have degenerate masses.

 \item[Fig. 4:] Momentum dependence of the three wave functions $|G_{f1}|$ 
                corresponding to the eigenvalues $\m^f$ shown in
                fig.~3, evaluated at the
                waveguide, $t=1$, on a $2+1$ dimensional lattice with
                $L=16$ and $L_s=50$ and $m_0=1.1$. The symbols
                correspond to those of fig. 3.

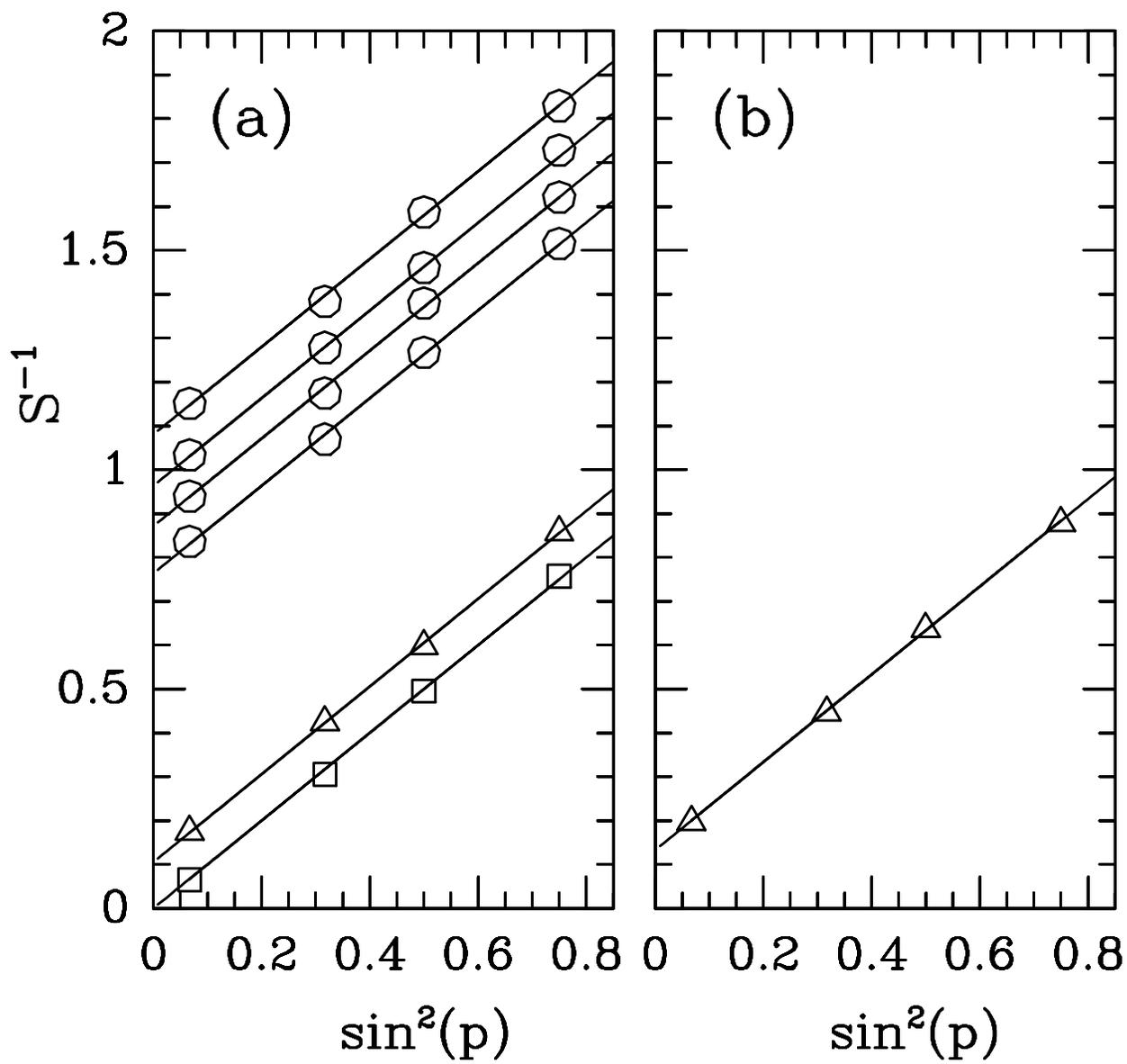
 \item[Fig. 5:] Inverse propagator $S^{-1}(p)$ measured at
                $s=1$, $s_0-1$, $s_0$, $L_s/2+1$, $s_0'$ and $s_0'+1$,
                for $y=0.5$ and $\k=0.5$ on a $12^2 26$ lattice with
                $m_0=1.1$.
                We show averaged components as explained in the text,
                errorbars are smaller that the symbols.
                For free naive fermions the fits with Ansatz 
                $S^{-1}(p)=(m_F^2 + \sum_\m\sin^2 p_\mu)/Z_F$
                (solid lines) would be exact.
                Figure a is for the full model which has the domain wall
                zeromode (squares), the light waveguide mode
                (triangles) and many heavy modes (circles); figure b
                is for the reduced model, which only contains the 
                waveguide mode.

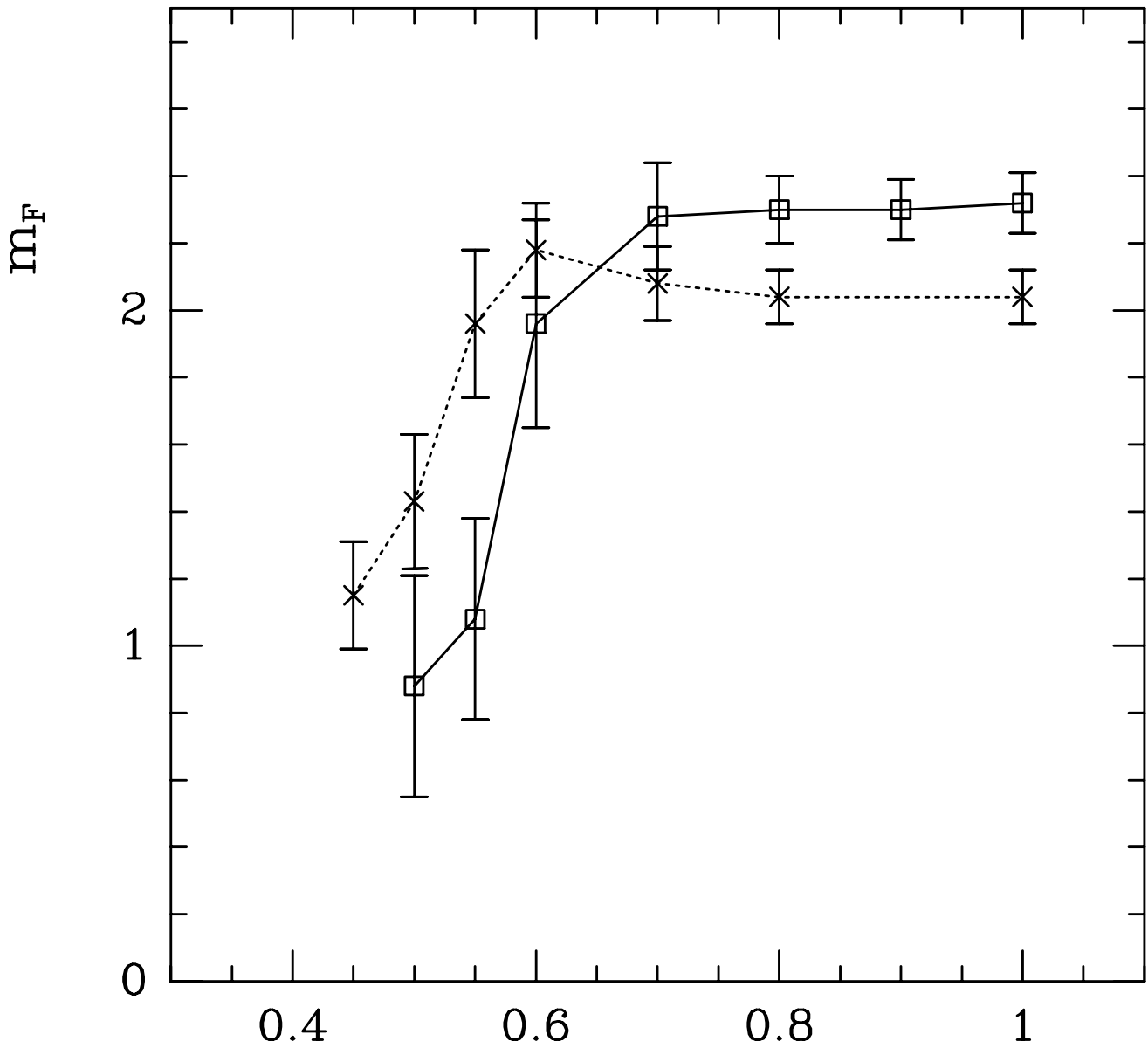
 \item[Fig. 6:] The $\k$ dependence of the waveguide fermion mass
                at strong coupling $y=2$,  on a $12^2 26$ lattice with
                $m_0=1.1$.
	        The boxes (crosses) are for the full (reduced) model.

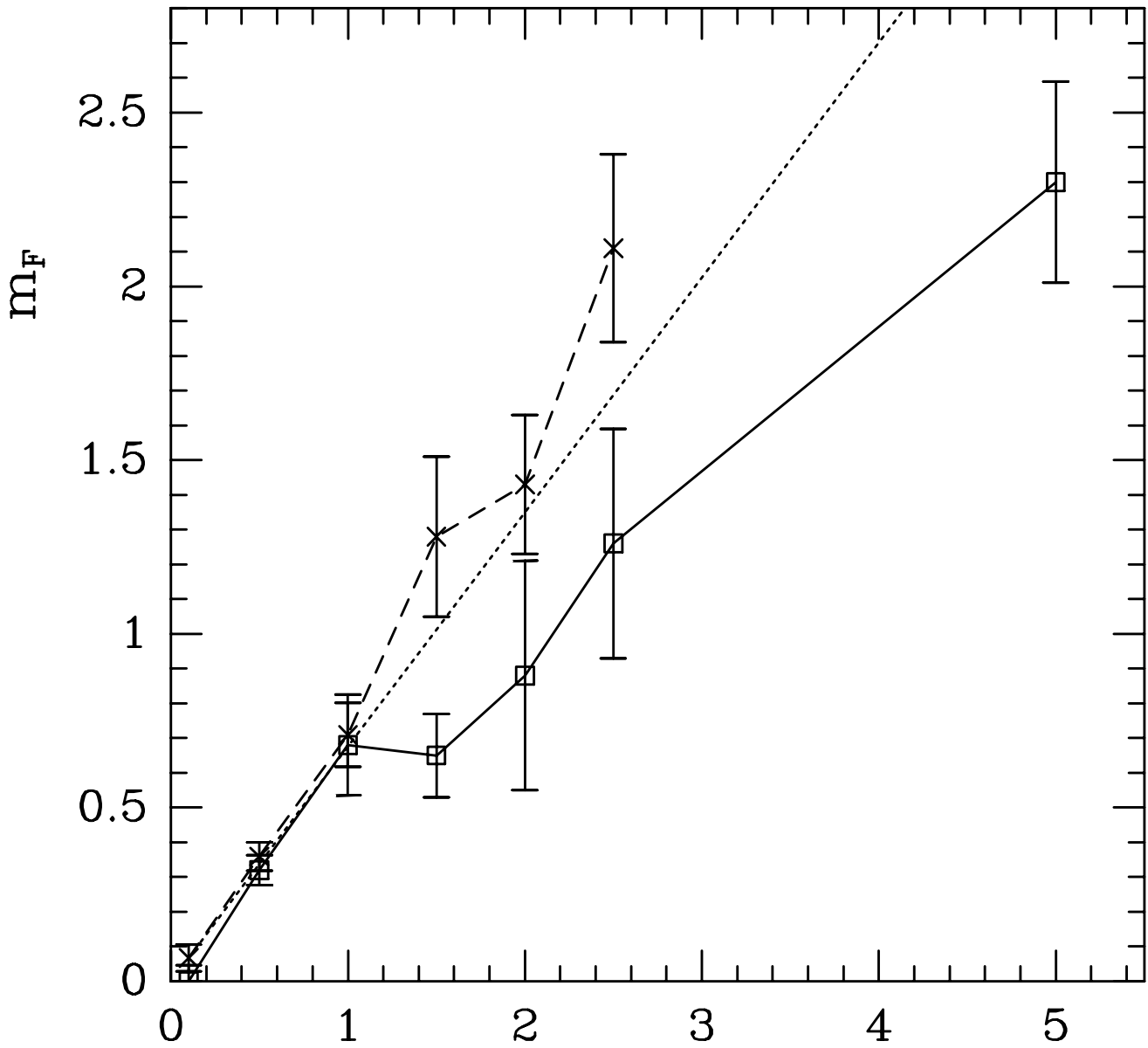
 \item[Fig. 7:] The $y$ dependence of the waveguide fermion mass
	        near the phase transition at $\k=0.5$,
                on a $12^2 26$ lattice with $m_0=1.1$.
	        The boxes (crosses) are for the full (reduced) model.

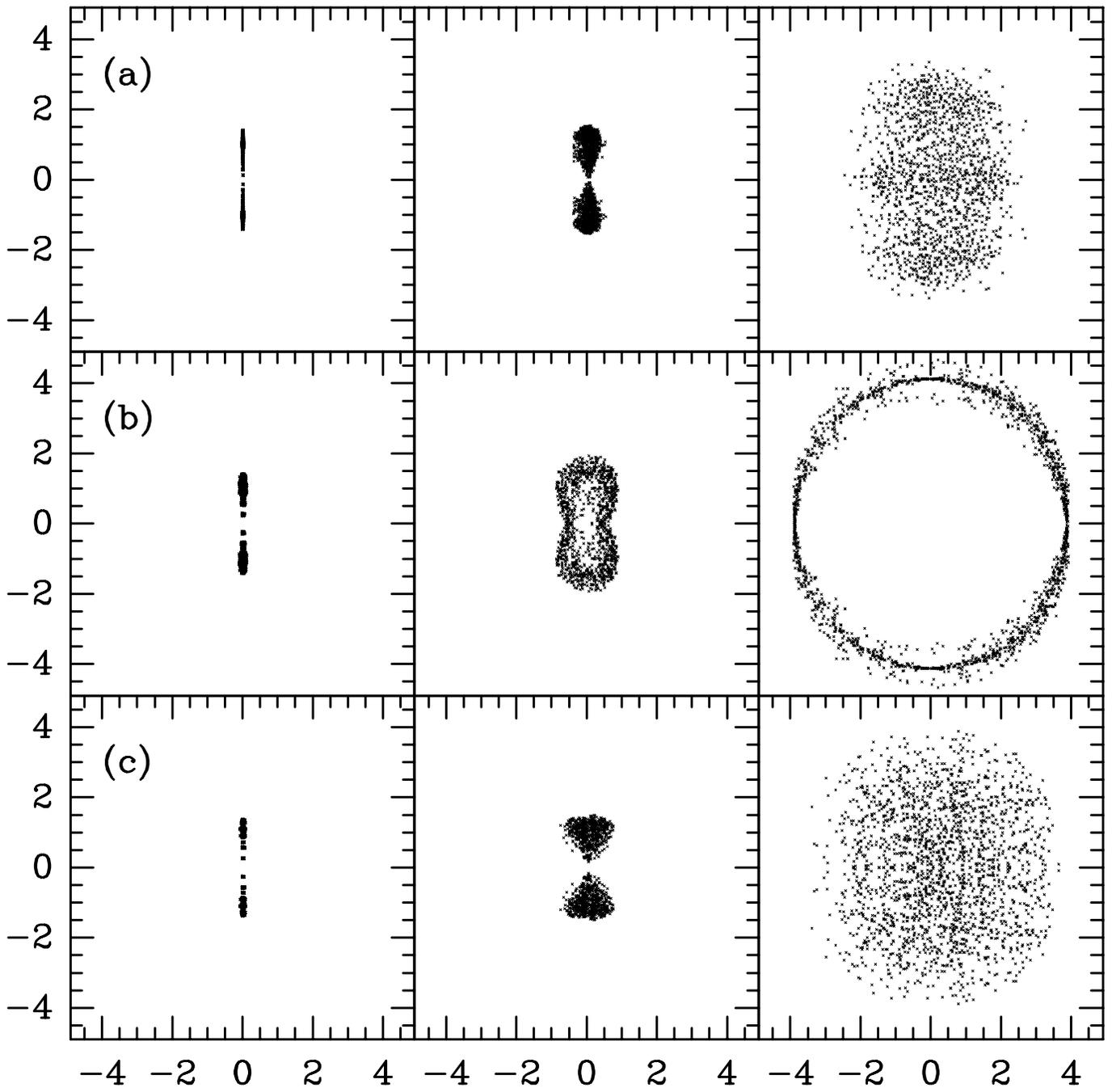
 \item[Fig. 8:] Eigenvalue spectra for the reduced domain wall fermion
                model (figures a), the reference Yukawa model with
                local (figures b) and hypercubical (figures c) coupling.
	        The left, middle and right figures are for $y=0.2$, $1.0$
                and $4.0$ respectively. 
                The lattice size is $L^2=12^2$ and $\k=0.1$.

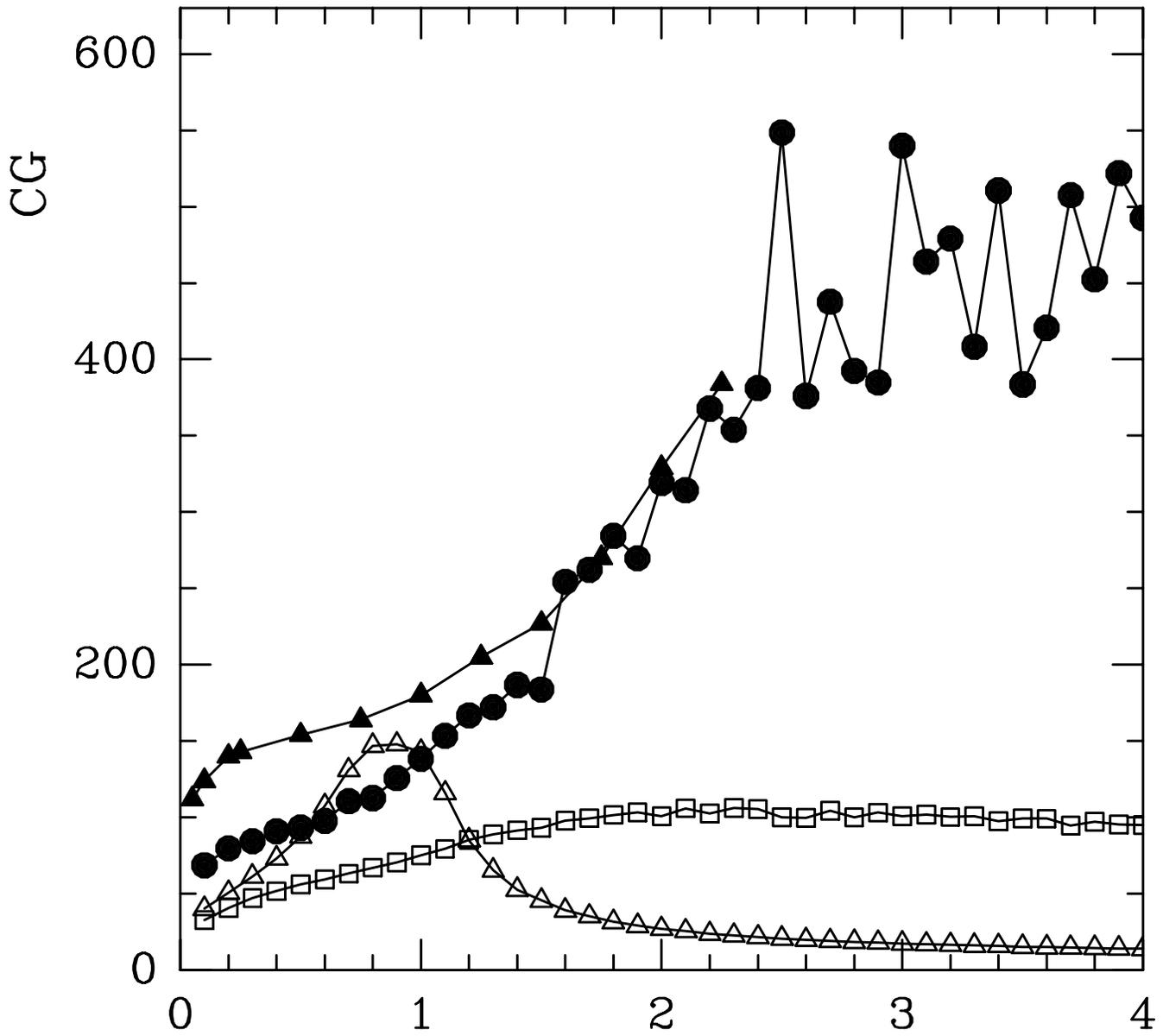
 \item[Fig. 9:] Number of conjugate gradient iterations to reduce the
                residual to $<10^{-6}$. The full symbols are for
                the quenched (circles) and unquenched (triangles)
                domain wall fermion model on a $12^2 26$ lattice. 
                The open symbols are for
                the Yukawa models defined in eq. \protect{(\ref{YMODEL})}
                with local (triangles) and hypercubical (boxes) Yukawa
                interaction, on a $12^2$ lattice with $\k=0.1$.

\end{description}
                
\end{document}